\documentclass[useAMS,usenatbib]{mn2e}
\usepackage{setspace}
\usepackage{amsmath}
\usepackage{amssymb}
\usepackage{graphicx}
\usepackage{multicol}
\usepackage[margin=2cm]{caption}
\usepackage{natbib}
\usepackage[T1]{fontenc}
\usepackage{aecompl}
\newcommand{\eq}[1]{\begin{equation}\begin{split}#1\end{split}\end{equation}}
\newcommand{\ea}[1]{\begin{align*}#1\end{align*}}
\newcommand{\eal}[1]{\begin{align}#1\end{align}}

\def\be{\begin{equation}}
\def\ee{\end{equation}}
\def\ba{\begin{eqnarray}}
\def\ea{\end{eqnarray}}
\def\go{\mathrel{\raise.3ex\hbox{$>$}\mkern-14mu
             \lower0.6ex\hbox{$\sim$}}}
\def\lo{\mathrel{\raise.3ex\hbox{$<$}\mkern-14mu
             \lower0.6ex\hbox{$\sim$}}}

\def\hatn{{\hat{\bf n}}}
\def\hatr{{\hat{\bf r}}}
\def\hate{{\hat{\bf e}}}

\begin{document}

\title[Evection Resonance of Planetary Systems]
{Disruption of Planetary Orbits Through Evection Resonance with an
  External Companion: Circumbinary Planets and Multiplanet Systems}
\author[Xu and Lai ]{Wenrui Xu and Dong Lai\\
Cornell Center for Astrophysics and Planetary Science,
Department of Astronomy, Cornell University, Ithaca, NY 14853, USA}

\maketitle

\begin{abstract}
Planets around binary stars and those in multiplanet systems may
experience resonant eccentricity excitation and disruption due to
perturbations from a distant stellar companion.  This ``evection                              
resonance'' occurs when the apsidal precession frequency of the
planet, driven by the quadrupole associated with the inner binary or the
other planets, matches the orbital frequency of the external
companion.  We develop an analytic theory to study the effects of
evection resonance on circumbinary planets and multiplanet systems. We
derive the general conditions for effective eccentricity excitation or
resonance capture of the planet as the system undergoes long-term
evolution.  Applying to circumbinary planets, we show that inward
planet migration may lead to eccentricty growth due to evection
resonance with an external perturber, and planets around shrinking
binaries may not survive the resonant eccentricity growth. On the
other hand, significant eccentricity excitation in multiplanet systems
occurs in limited parameter space of planet and binary semimajor axes,
and requires the planetary migration to be sufficiently slow.
\end{abstract}


\section{Introduction}

Evection resonance is a secular-orbital resonance between the apsidal
precession of a planet or satellite due to the quadrupole moment of a
central body and the orbital motion of a distant perturber. This
resonance was first studied in the context of the Moon-Earth-Sun
system by \cite{TW98}, who showed that the resonance between the
Moon's orbital precession due to the Earth's oblateness and the
periodic perturbation of the Sun could give the Moon a significant
eccentricity in its early evolution, which later produce the
misalignment of the Moon's orbit. Recently, \cite{SBA16}
studied the evolution of exomoons during planetary migration, and
showed that evection resonance could induce eccentricity growth of
exomoons, leading to their collisions with the host planet.

A recent study by \cite{TS15} examined the effect of evection
resonance in multiplanet systems perturbed by distant binary
companions.  For such a system, the evection resonance occurs between
one of the precession modes of the planets, which is similar to the
apsidal precession the outer planet due to the quadrupole moment from
the inner massive planets, and the orbital motion of the binary. It
was showed that planetary migration can trap the outer planet into an
evection resonance, resulting in significant planetray eccentricity
excitation and possibly disruption.  \cite{TS15} suggested that such
resonant driving of eccentricity by binary may have altered the
architecture of many multiplanet systems and weaken the multiplanet
occurrence rate in wide binaries.

Evection resonance may also play role in circumbinary planet systems.
About 10 binary star systems harboring transiting circumbinary
planets have been discovered by the \textit{Kepler} mission 
(e.g. \citealt{Doyle11}; \citealt{Kostov15}).
However, no transiting circumbinary planet has been found around
compact stellar binaries with periods $\lesssim 7$ days, despite 
the large number of such compact eclipsing binaries
in the {\it Kepler} sample. 
It is generally believed that close binaries (with periods $\lesssim
5$ days) are not primordial, but have formed at a wider separation and
subsequently shrunk via Lidov-Kozai (LK) cycles with tidal friction induced
by an inclined tertiary companion \citep{FT07}.  Several papers have
examined the dynamics of ``binary + planet'' systems in the presence of
distant stellar companions (\citealt{ML15}; \citealt{Martin15}; \citealt{Hamers16}).
A sufficiently massive planet can suppress the shrinkage of the inner
binary orbit by disrupting the LK cycles, thus explaining the
lack of planet-hosting short-period binaries. Alternatively, a
low-mass circumbinary planet does not affect the LK cycles of
the inner binary, but becomes misaligned with the shrinking binary or
becomes unstable and destructed during the binary shrinkage \citep{ML15}.  However, in all these studies, the effect of evection
resonance was not considered.  As the inner binary undergoes LK oscillations
and orbital decay, the apsidal precession of the planet may be resonant with the
orbital motion of the tertiary companion. This can excite planetary eccentricity 
and may lead to the destruction of the planet.

In this paper we develop an analytic theory to study the effects of
evection resonance on circumbinary planets and multiplanet systems
induced by external stellar companions.  In particular, we derive the
general conditions for which appreciable eccentricity excitation
associated with resonance passage or resonance capture can be
achieved. These conditions and the expression for the maximum eccentricity can be
applied to a variety of situations. 

Our paper is structured as follows.  In Section 2 we summarize the
general theory of evection resonance for a circumbinary planet under
the perturbation of a tertiary companion. We use the Hamiltonian
approach, ignoring dissipative effects and long term evolution of the
system (e.g., associated with planet migration or LK cycles of the
inner binary). We obtain a one-parameter nondimensionalized
Hamiltonian and use it to calculate the width and libration timescale
of the resonance.  In Section 3 we study the passage through resonance
due to the long term evolution of the system by considering the
evolution of the parameter of the nondimensionalized Hamiltonian.  We
obtain the criteria for resonant trapping, and estimate the magnitude
of eccentricity excitation. In Section 4 we apply the results in
Section 3 to realistic long-term evolutions of hierarchical triple
systems hosting a circumbinary planet, including planet migration and
the LK oscillations and orbital decay of the inner binary.
In Section 5 we adapt the theory developed in Sections 2-3 to multiplanet systems
with external perturbers, and examine the condition for planetary eccentricity
excitation due to evection resonance.  We summarize our results in Section 6.

\section{Evection Resonance of a Circumbinary Planet with an External Perturber}

\subsection{Setup and Perturbation Potential}

Consider a binary with masses $M_1,M_2$ and semimajor axis $a_b$
orbited by a planet of mass $m_p$ and semimajor axis $a$ [relative to
  the center of mass (CM) of the binary]. The binary is in a
hierarchical triple system, with an extrenal perturber of mass $M_B$
and semimajor axis $a_B$ (relative to the CM of the inner binary).
We also define $M_b=M_1+M_2$ and $\mu_b=M_1M_2/M_b$ as the total mass
and reduced mass of the inner binary, and $M_{\rm tot}=M_b+M_B$ as the
total mass of the stellar triple. Throughout the paper we
consider $m_p\ll M_1,M_2,M_B$ and $a_b\ll a\ll a_B$.  For simplicity,
we assume that the outer binary is circular ($e_B=0$), but the inner
binary and the planet can have general eccentricities and
inclinations. Also, in this section and Section 3 we ignore the
evolution of the inner binary driven by $M_B$, because the timescale
for this evolution is much longer than the timescale for the evolution
of the planet's orbit.

The planet experiences perturbations from both the inner and outer
binaries. To the quadrupole order, the perturbing potential acting on
the planet can be written as
\eq{
\Phi = \Phi_b+\Phi_B,
}
where $\Phi_b$ is the potential from the inner binary (double-averaged over the orbits
the planet and inner binary),
\eq{\label{eq:Phib}
\Phi_b =& \frac{\Phi_{b0}}{8(1-e^2)^{3/2}}\left[1-6e_b^2-3(1-e_b^2)
(\hatn_b\cdot\hatn)^2 \right.\\
&\left. +15\,e_b^2\,(\hate_b\cdot\hatn)^2
\right],}
with
\eq{
\Phi_{b0} = \frac{\mathcal G \mu_b a_b^2}{a^3},}
and $\Phi_B$ is the potential from the outer binary (averaged over the orbit of the planet),
\eq{\label{eq:PhiB}
\Phi_B  =& \frac{\Phi_{B0}}{4}\left[-1+6e^2+3(1-e^2)(\hatn\cdot\hatr_B)^2\right.\\
&\left. -15\,e^2(\hate\cdot\hatr_B)^2\right],}
with
\eq{\Phi_{B0}= \frac{\mathcal GM_Ba^2}{a_B^3}.}
In Eqs.~(\ref{eq:Phib}) and (\ref{eq:PhiB}),
$\hate$ is the unit vector in the direction of periapsis and $\hatn$ is the unit vector 
normal to the orbital plane of the planet, while $\hate_b$ and $\hatn_b$ are the corresponding
unit vectors for the inner binary; $\hatr_B$ is the unit vector in the direction of 
$M_B$. The derivation of these potentials can be found in, e.g., \cite{TTN09} and
\cite{TY14}. Note that when $e_b\neq 0$ and $M_1\neq M_2$, the inner binary also exerts
an octupole potential on the planet (e.g. Liu et al.~2015), but 
we ignore it here because it tends to be small for $M_1\sim M_2$. The outer binary has a 
zero octupole potential because it has zero eccentricity.

The ratio $\epsilon\equiv \Phi_{B0}/\Phi_{b0}$
characterizes the relative strengths of the two potentials. 
Throughout this paper, we consider the regime where
\eq{\label{epsilon_def} 
\epsilon\equiv \frac{\Phi_{B0}}{\Phi_{b0}}={M_B\over\mu_b}{a^5\over a_b^2a_B^3}\ll 1,}
i.e., the potential from the inner binary dominates over the outer potential. In another word,
the planet lies inside the Laplace radius $a_L$, i.e.,
\eq{
a\ll a_L\equiv \left({\mu_{b}\over M_B}\right)^{1/5}\left(a_b^2a_B^3\right)^{1/5}.}

To express the potential in terms of various orbital elements, we set up a Cartesian
cooridnate system with the $z$-axis along $\hatn_b$, and the $x$-axis along the direction
of $\hatn_b\times\hatn_B$ (so the longitude of ascending node of the outer binary 
is $\Omega_B=0$). In this coordinate system, we have
\eal{
\hat{\mathbf n} =& \left(\begin{array}{l}\sin\Omega\sin I\\-\cos\Omega\sin I\\\cos I\end{array}\right),\\
\hat{\mathbf e} =&  \left(\begin{array}{l}\cos\Omega\cos\omega-\sin\Omega\sin\omega\cos I\\\cos\Omega\sin\omega\cos I+\sin\Omega\cos\omega\\ \sin\omega\sin I\end{array}\right),\\
\hat{\mathbf r}_B =& \left(\begin{array}{l}\cos\lambda_B\\\sin\lambda_B\cos I_B\\\sin\lambda_B\sin I_B\end{array}\right),
}
where $I$, $I_B$ are the inclination angles of the planetary and outer
binary orbits relative to the inner binary orbit (i.e., $\cos I\equiv
\hatn\cdot\hatn_b$ and $\cos I_B\equiv \hatn\cdot\hatn_B$),
$\Omega$ and $\omega$ are the longitude of ascending node and the
argument of pericenter of the planet's orbit, and $\lambda_B$ is the mean longitude of the
external perturber $M_B$. The eccentricty unit vector of the inner binary is 
$\hate_b=(\cos\varpi_b,\sin\varpi_b,0)$, with a constant $\varpi_b$
(the longitude of pericenter).

\subsection{Hamiltonian Near Resonance}

Since $\epsilon\ll 1$, the nodal precession of the planet induced by
$\Phi_b$ is much faster than that induced by $\Phi_B$, and the
planet's orbit is strongly coupled to the inner binary. Thus, if the initial inclination angle
$I$ is zero, the planet will stay aligned with the inner binary even in
the presence of perturbation from the outer binary \citep{ML15}. 
In the following analysis, we assume $I=0$ at all times; this greatly simplifies
the calculation without much loss of generality. In this case, the
apsidal precession rate (for the longitude of pericenter $\varpi=\Omega+\omega$)
of the planet induced by the inner binary is given by
\eq{\label{eq:dotomega}
\dot\varpi={3(2+3e_b^2)\over 8(1-e^2)^2}\,\frac{\Phi_{b0}}{a^2n}
\simeq {3(2+3e_b^2)\over 8}\,\frac{\Phi_{b0}}{a^2n},}
where $n=\sqrt{\mathcal G M_{b}/a^3}$ (with $M_{b}=M_1+M_2$) 
is the mean motion of the planet, and the second equality assumes $e\ll 1$.

Consider a planet near the ``evection resonance'', where the precession rate of
$\varpi$ is close to the outer binary's orbital frequency $n_B=\sqrt{\mathcal GM_{\rm tot}/a_B^3}$
(where $M_{\rm tot}=M_{b}+M_B$), i.e.,
\be
\dot\varpi\simeq n_B=\dot\lambda_B,
\ee
or equivalently
\be
{{\dot\varpi}^2\over n_B^2}=
\left[{3(2+3e_b^2)\over 8}\right]^2\left({\mu_{b}^2\over M_{b}M_{\rm tot}}
\right)\left({a_b^4a_B^3\over a^7}\right)
\simeq 1.
\label{eq:res2}\ee
Physically, the resonance occurs when the planet's eccentricity 
vector librates around a fixed angle with respect to the outer binary.
Avergaing out the fast-varying angles $\varpi$ and $\lambda_B$, we find 
(for $I=0$)
\eq{
\Phi =& -\Phi_{b0}\left[\frac{2+3e_b^2}{8(1-e^2)^{3/2}} 
+{\epsilon\over 16}\,\Bigl[(6+9e^2)\cos^2\! I_B-(2+3e^2)\Bigr] \right.
\\&\left.  +{15\,\epsilon \over 32}\, (1+\cos I_B)^2\, e^2 \cos(2\varpi-2\lambda_B)
\right].}
We can nondimensionalize the Hamiltonian in the canonical coordinate and momentum
(the modified Delaunay variables)
\eq{
\gamma = -\varpi,\quad \Gamma = 1-(1-e^2)^{1/2}\simeq {e^2\over 2}.
}
The dimensionless time is 
\be
\hat t \equiv t\,\frac{\Phi_{b0}}{n a^2},
\ee
and the dimensionless Hamiltonian (for $e\ll 1$) is 
\eq{
\hat H\equiv\frac{H}{\Phi_{b0}}
=( A+\epsilon B)\Gamma + C\Gamma^2 +\epsilon D\Gamma\cos(2\gamma+2\lambda_B),}
where we have dropped a non-essential constant.
The dimensionless 
constants $A,B,C,D$ (all of order unity) are given by 
\eal{
A =& -\frac{3}{8}(2+3e_b^2),\\
B = &\frac{3}{8}(1-3\cos^2\!I_B),\\
C = & -\frac{3}{4}(2+3e_b^2),\\
D = &-\frac{15}{16}(1+\cos I_B)^2.
}

To further simplify the Hamiltonian, we make another canonical transformation and rescale the 
Hamiltonian:
\eal{
\theta =& -2\gamma - 2\lambda_B=2\varpi-2\lambda_B,\\
\Theta =& \frac{C}{\epsilon D}\Gamma,\\
K =& \frac{C}{\epsilon^2D^2}\hat H+\frac{n_B n a^2}{\epsilon D\Phi_{b0}}\Theta,\\
\label{time_scaling}
\tau =& -2\epsilon D\hat t = -\frac{2 \epsilon D\Phi_{b0}}{n a^2}\,t.
}
The new (dimensionless) Hamiltonian is then
\eq{\label{nond_hamiltonian}
K = \eta \Theta + \Theta^2 + \Theta\cos\theta,
}
where $\eta$ is a constant parameter given by 
\eq{\label{eta_def}
\eta = \frac{A+\epsilon B+(n_Bna^2/\Phi_{b0})}{\epsilon D}.
}
Using Eq.~(\ref{eq:dotomega}) we find
\eq{\label{eq:eta2}
\eta = {2(2+3e_b^2)\over 5(1+\cos I_B)^2}\,\,{1\over\epsilon}
\left[1-{n_B\over\dot\varpi(e=0)}-\epsilon\,\left({1-3\cos^2\!I_B\over 2+3e_b^2}\right)\right].}
Note that the dimensionless time $\tau$ for the Hamiltonian (\ref{nond_hamiltonian})
can be written as
\be
\tau=2|D|\, {t\over T_K},
\label{eq:tau}\ee
where 
\be
T_K^{-1}\equiv {\Phi_{B0}\over na^2}={M_B\over M_{b}}\left({a\over a_B}\right)^3 n
\label{eq:tk}\ee
characterizes the precession rate (Kozai rate) of the planet driven by the
extrenal companion. The dynamical variable $\Theta$ is related to the eccentricity
($e\ll 1$) by
\eq{
\Theta \simeq {C\over 2D} {e^2\over\epsilon}.
}

The Hamiltonian \eqref{nond_hamiltonian} is the same as that
describing the dynamics of planets near the second-order mean-motion
resonance \citep{MD99}.  Thus we should expect that the resonance
structure to be the same.

\subsection{Structure of Resonance}

\begin{figure}
\centering
\includegraphics[width=8.3cm]{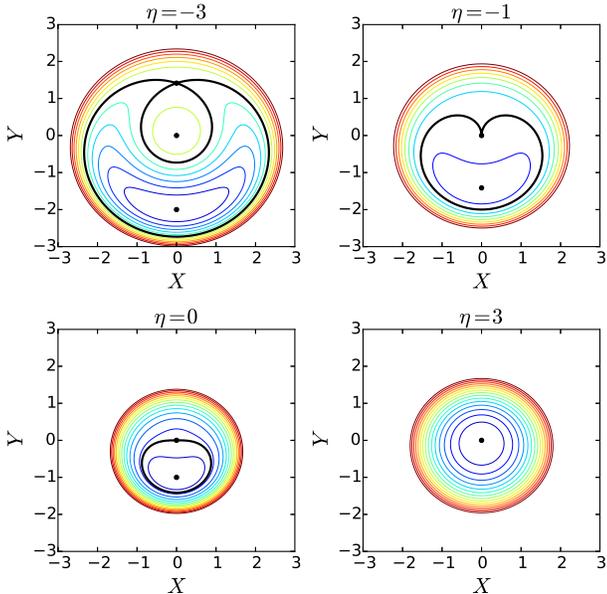}
\caption{Level curves of the Hamiltonian for different $\eta$
  values. The black dots and black lines mark the fixed points and separatrices. 
  Two bifurcations take place at $\eta=-1$ and $\eta=1$. The bifurcation
  at $\eta=-1$ is shown. For $\eta<1$, there exists a libration zone;
  for $-1<\eta<1$, the trajectory for an initially circular orbit lies close
  to the separatrix and can be excited to nontrivial eccentricities.}
\label{phase_space_structure}
\end{figure}

Next we can examine the structure of the resonance by studying the
phase space topology of the system. It is convenient to use the conjugate
(Poincare) variables
\eq{
X = \sqrt{2\Theta}\sin\theta,~~~Y = \sqrt{2\Theta}\cos\theta.
}
The Hamiltonian becomes
\eq{
K = \frac{\eta}{2} (X^2+Y^2) + \frac{1}{4}(X^2+Y^2)^2 + \frac{1}{2}Y\sqrt{X^2+Y^2}.
}
The level curves of this Hamiltonian for different $\eta$ values are
shown in Fig.~\ref{phase_space_structure}.
Note that $\sqrt{X^2+Y^2}= \sqrt{2\Theta}\simeq 
e\sqrt{C/\epsilon D}$, the radius from the origin to the 
trajectory measures $e/\sqrt{\epsilon}$.

The fixed points of the Hamiltonian follow from $\partial K/\partial
X=\partial K/\partial Y=0$, and are located at $X=0$ and $Y=0$,
$Y=\sqrt{-1-\eta}$ or $Y=-\sqrt{1-\eta}$.  Thus, for $\eta\ge 1$,
there is one stable fixed point at $(X,Y)=(0,0)$; for $\eta<1$, there
are two fixed points at $(X,Y)=(0,0)$ and $(0,-\sqrt{1-\eta})$; for
$\eta <-1$, there is an additional (saddle) point at
$(0,\sqrt{-1-\eta})$.

We see that for $-1\leq\eta<1$, the phase-space trajectory of 
an initially near-circular orbit (which lies near the origin of the
$XY$ space) is close to the separatrix, giving rise to 
nontrivial eccentricity excitation that depends weakly on the initial eccentricity.
For a given $\eta\in [-1,1)$, the maximum $\Theta$ that can be achieved
(corresponding to the minimum $Y$ on the separatrix) is 
\eq{
\Theta_{\text{max}}(\eta) = 1-\eta.
}
This gives
\eq{
e_{\text{max}}(\eta) = \sqrt{\frac{2\epsilon D}{C}(1-\eta)}\qquad {\rm with}~\eta\in [-1,1)
}
The ``limiting eccentricity'' (i.e. the maximum value of $e_{\text{max}}$ for all $\eta$) occurs
at $\eta=-1$ and is given by 
\eq{\label{eq:elim}
e_{\rm lim} =e_{\rm max}(\eta=-1)= 2\sqrt{\frac{\epsilon D}{C}}.
}

It is useful to estimate the periods of trajectories in the libration zone.
These periods are generally of order unity (in terms of the dimensionless tie $\tau$),
but can become larger for $\eta$ close to $\pm 1$.
The equation of motion is given by
\eal{\label{eom_ham_1}
\frac{d\theta}{d\tau} &= \frac{\partial K}{\partial \Theta} =  \eta + 2\Theta + \cos\theta,\\
\label{eom_ham_2}
\frac{d\Theta}{d\tau} &= -\frac{\partial K}{\partial \theta} = \Theta\sin\theta.
}
When the trajectory has a small initial eccentricity, or
$\Theta_0\equiv\Theta(t=0)\ll 1$, the period approximately corresponds 
to the time required to reach $\Theta\sim 1$. 
The system spends most time near the turning point of
$\theta$-libration, where (for small $\Theta$) we have
$\cos\theta\simeq-\eta$ and $d\ln\Theta/d\tau =
\sin\theta\sim\sqrt{1-\eta^2}$.
Thus the (dimensionless) libration period for a trajectory with small
$\Theta_0$ is
\eq{\label{dimless_period}
\tau_{\rm lib} \sim \frac{|\ln \Theta_0|}{\sqrt{1-\eta^2}}.
}

Also of interest is the trajectory doing small-amplitude oscillation
around the stable fixed point $(X,Y)=(0,-\sqrt{1-\eta})$ 
[corresponding to $\theta=\pi$, $\Theta=(1-\eta)/2$] in the libration zone.
Expanding Eqs.~(\ref{eom_ham_1})-(\ref{eom_ham_2}) around the fixed point
(e.g. $\theta=\pi+\Delta\theta$), we find $d^2\Delta\theta/d\tau^2\simeq -(1-\eta)
\Delta\theta$. The period is then
\eq{\label{dimless_period2}
\tau_{\rm lib}\simeq \frac{2\pi}{\sqrt{1-\eta}},
}
which is of order unity unless $\eta$ approaches $1$.

\section{Resonant Passage and Capture}

Having studied the dynamics of our planet-in-binaries
system near the evection resonance
at a constant $\eta$, we now examine the evolution of the planet's eccentricity
as the parameter $\eta$ changes slowly. Such gradual change of $\eta$ 
can be facilitated by the migration of the planet or the evolution of the inner
binary -- these specific applications will be discussed in Section 4.
The slow evolution of $\eta$ can produce nontrivial 
excitation of the planet's eccentricity, as the system is driven across the resonance
and sometimes gets trapped in resonance.  Since equation
\eqref{nond_hamiltonian} has the same form as the Hamiltonian for the
second-order mean motion resonance, many aspects of the resonance
capture problem have been studied before (see \citealt{MD99},
\citealt{PEALE86} and \citealt{BG84}). These studies, however, mainly
focused on the case where the orbit has a relatively large initial
eccentricity (i.e. the initial $\Theta\gtrsim 1$) and the
evolution of $\eta$ is infinitely slow.

In the following, we assume that the system is initially out of
resonance with a small eccentricity $e_0$.
We focus on the regime where $\epsilon$ is large enough so that
$\Theta_0=|C/2D|(e_0^2/\epsilon)\ll 1$, which has not been covered by pervious studies. 
We also assume that $d\eta/d\tau$ is constant. We consider a range of values for
$|d\eta/d\tau|$, from $\ll 1$ to $\sim 1$ [note that the libration rate
  of the resonance is of order unity; see
  Eqs.~(\ref{dimless_period})-(\ref{dimless_period2})].  The initial
$\eta_0\equiv \eta (t=0)$ is chosen such that the system passes
through resonance at some later times.

\subsection{Increasing $\eta$: Eccentricity Excitation Without Trapping}

\begin{figure}
\centering
\includegraphics[width=8.3cm]{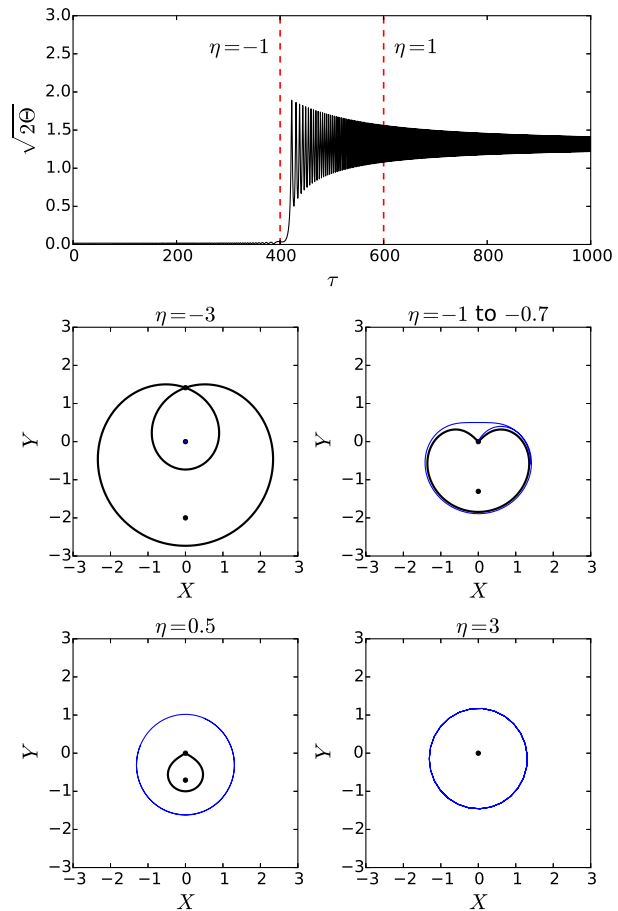}
\caption{Resonant passage for $d\eta/d\tau = 10^{-2}$ and $\Theta_0=10^{-4}$. 
The top panel shows the dimensionless
  eccentricity ($\sqrt{2\Theta}\sim e/\sqrt{\epsilon}$) evolution as a function of time $\tau$, 
  with the boundaries of the resonance ($\eta=\pm1$) marked by
  red dashed lines. The bottom four panels show the phase-space trajectories 
  for periods of time near certain $\eta$ values or within a range of $\eta$. The blue
  curves are trajectories of the system, and the black curves and dots mark
  the separatrices and fixed points. The planet is initially at a
  low-eccentricity state ($\eta=-3$); it experiences a non-adiabatic ``jump''
  near $\eta = -1$, and subsequent adiabatic evolution ($\eta=0.5$ and $\eta=3$)
  during which the phase-space area of the trajectory is conserved.}
\label{resonant_passage}
\end{figure}

Consider a system initially at $\eta_0<-1$ and  $\Theta_0\ll 1$.
As the parameter $\eta$ gradually increases, the system passes through 
the resonance and may experience significant eccentricity growth.
Figure \ref{resonant_passage} depicts an example.

First consider the case when $|d\eta/d\tau| \ll 1$. Away from the sepratrix, the
dynamical (libration/circulation) time of the system is $\tau_{\rm lib}\sim 1$.
So when $|d\eta/d\tau|\ll 1$, the theory of adiabatic invariance implies 
that the phase-space area covered by the trajectory is conserved, i.e.
\be
{\cal A}=\oint\, Y\,dX=\oint\,\Theta\,d\theta={\rm constant}.
\ee
Thus, starting from $\eta_0<-1$ (where the system lies in the inner circulating zone),
the planet maintains at its small eccentricty for
$\eta<-1$. Then, near $\eta=-1$, the trajectory encounters the separatrix and experiences
a jump in $\cal A$ (and eccentricity). 
 After passing through $\eta=-1$, the separatrix continues to shrink
and the system trajectory lies in the outer circulating zone, conserving its $\cal A$.
Clearly, in the limit $|d\eta/d\tau|\ll 1$, the final $\cal A$ is equal to the area of the
critical ($\eta=-1$) separatrix (the homoclinic orbit). Using Eq.~(\ref{nond_hamiltonian}), 
this area is given by 
\eq{
\mathcal{A}_{\text{max}} = \int_0^{2\pi}d\theta(1-\cos\theta) = 2\pi.}
The subscript ``max'' implies that 
the final planet eccentricity attains its maximum value when ${\cal A}_{\rm f}={\cal A}_{\rm max}$
and this is achieved in the limit of $d\eta/d\tau\rightarrow 0$ (and $\Theta_0\rightarrow 0$):
\be
\Theta_{\rm f,max}=1,\quad {\rm or}\quad e_{\rm f,max}=\sqrt{2\epsilon D\over C}.
\label{THfmax}
\ee
Note that $e_{\rm f,max}=e_{\rm lim}/\sqrt{2}$, where $e_{\rm lim}$ 
[see Eq.~(\ref{eq:elim})] is the maximum 
eccentricity the planet experieneces {\it during} the resonance passage.

\begin{figure}
\centering
\includegraphics[width=8.3cm]{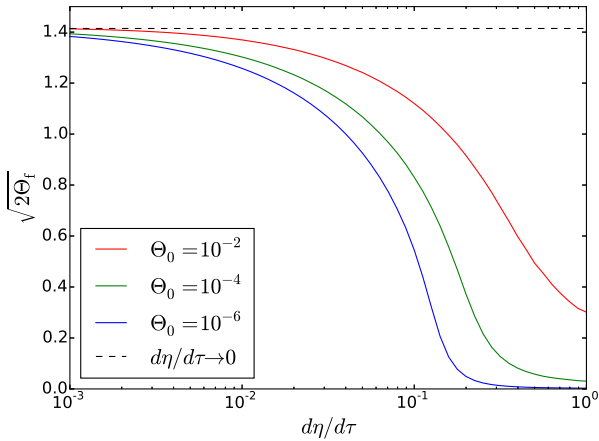}
\caption{Final eccentricity ($\sqrt{2\Theta_{\rm f}}\sim e_{\rm f}/\sqrt{\epsilon}$) 
  generated in a resonance passage as a
  function of $d\eta/d\tau$ for different values of initial eccentricity $\Theta_0$. The
  dashed line marks the theoretical value for the maximum eccentricity
  excitation (achieved for $d\eta/d\tau \to 0$), as calculated in
  Eq.~\eqref{THfmax}}
\label{Ecc_vs_deta}
\end{figure}

As $d\eta/d\tau$ increases, the final planet eccentricity
$e_{\rm f}$ after resonance passage becomes smaller than $e_{\rm f,max}$.
Figure~\ref{Ecc_vs_deta} shows our numerical results of $\sqrt{2\Theta_{\rm f}}
\sim e_{\rm f}/\sqrt{\epsilon}$ as a function of $d\eta/d\tau$ for several values of $\Theta_0$.
Clearly, the eccentricity excitation can be significantly reduced when $d\eta/d\tau$ 
is too large. The critical value, $(d\eta/d\tau)_c$, above which $e_{\rm f}$ becomes less
than $e_{\rm f,max}/2$, is of order $0.1$, but depends on the initial $\Theta_0$, and 
is smaller for smaller $\Theta_0$. This behavior can be 
qualitatively understood from Eq.~(\ref{dimless_period}), which shows that the libration
period for an initially low-eccentricity trajectory 
is proportional to $|\ln\Theta_0|$. Thus, to achieve significant eccentricity
excitation ($e_{\rm f}\go e_{\rm f,max}/2$) in a resonance passage, we require 
\be
\left|{d\eta\over d\tau}\right|\lo {1\over |\ln\Theta_0|}.
\label{conditon_on_init}
\ee

\subsection{Decreasing $\eta$: Resonance Trapping}

\begin{figure}
\centering
\includegraphics[width=8.3cm]{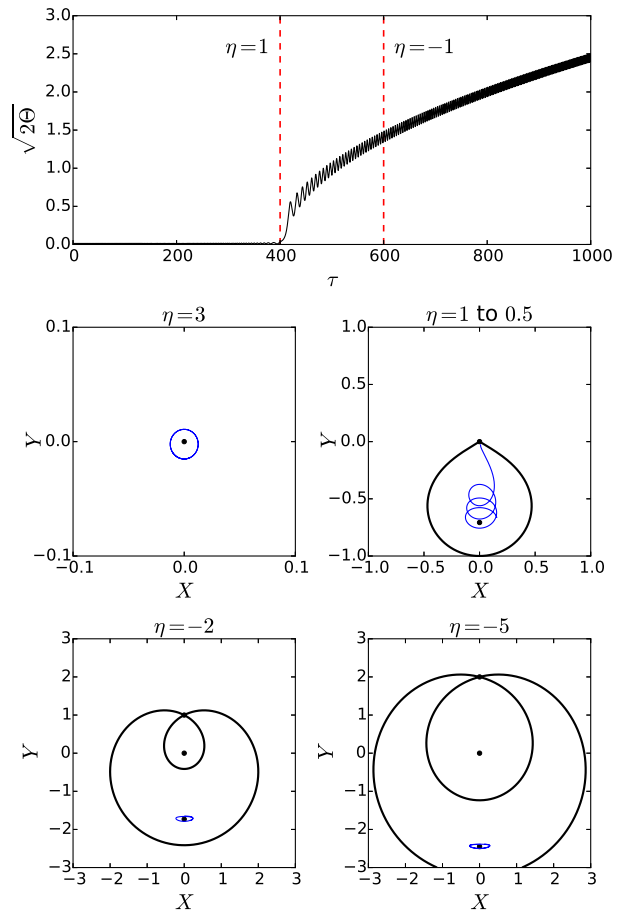}
\caption{Resonant trapping for $d\eta/d\tau = -10^{-2}$ and
  $\Theta_0=10^{-4}$ (cf.~Fig.~\ref{resonant_passage}).  The
  planet is initially in a low-eccentricity state ($\eta=3$).  At
  $\eta\le 1$, a libration (resonance) zone bifurcates from the
  origin, and the planet gets trapped in the resonance
  ($\eta=1$-$0.5$). As $\eta$ continues to decrease, the trajectory ``follows''
  the center (fixed point) of the resonance ($\eta=-2$ and $-5$) with ever
  increasing eccentricity.}
\label{Trap}
\end{figure}

\begin{figure}
\centering
\includegraphics[width=8.3cm]{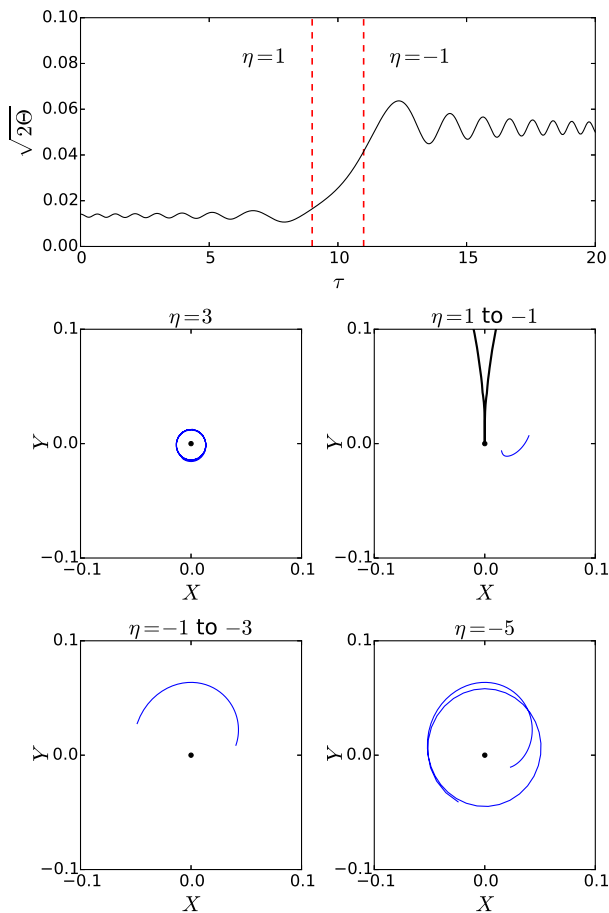}
\caption{Same as Fig.~\ref{Trap}, but with $d\eta/d\tau = -1$ and
  $\Theta_0 = 10^{-4}$. The axes for eccentricity evolution and phase
  space trajectories are zoomed-in to show the small eccentricity. 
During $1>\eta>-1$ the eccentricity grows as a result of the fast changing 
$\eta$. The system is not trapped into resonance; the
phase-space trajectory always circulates around the origin.}
\label{NotTrap}
\end{figure}

\begin{figure}
\centering
\includegraphics[width=8.3cm]{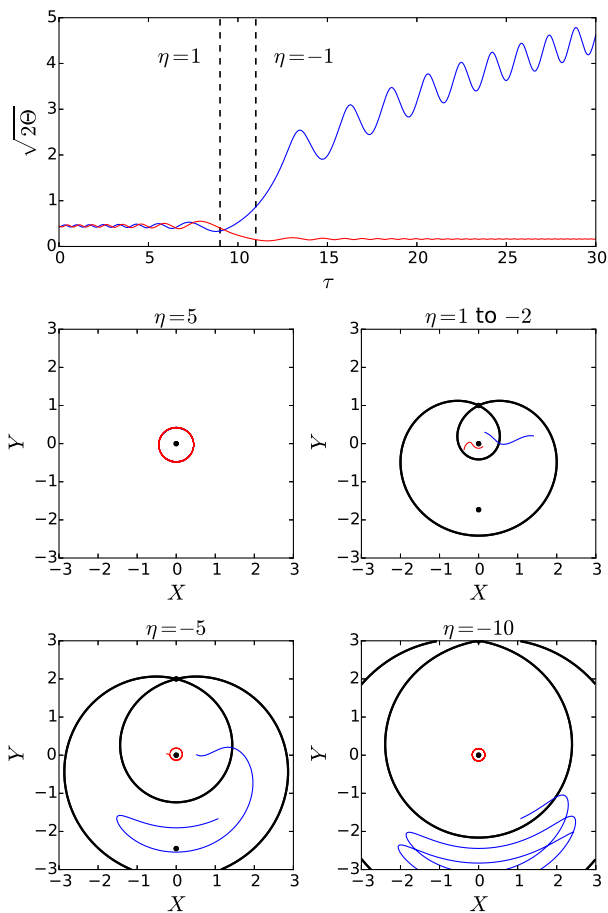}
\caption{Same as Fig.~\ref{Trap}, but with $d\eta/d\tau=-1$ and
  $\Theta_0=10^{-1}$ (a large $\Theta_0$ is chosen to better visualize
  the difference between the two cases). The blue and red curves
  correspond to two systems that differ only by $\pi/2$ in the initial
  phase ($\theta$). One of the systems is trapped (blue), while the
  other is not (red).  Trapping should be considered probabilistic in
  this regime, since the phase for a realistic system is arbitrary.}
\label{ProbTrap}
\end{figure}

\begin{figure}
\centering
\includegraphics[width=8.3cm]{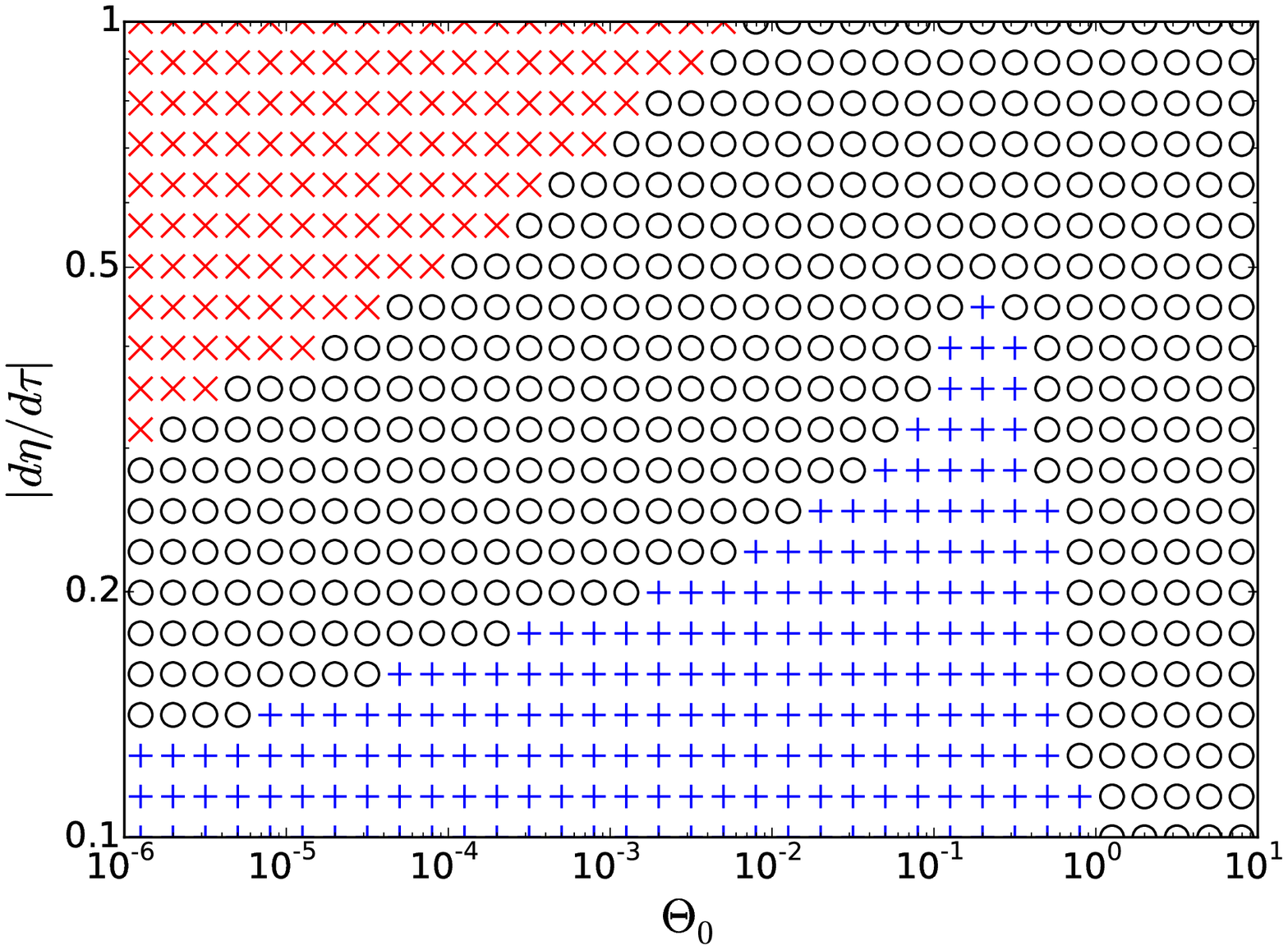}
\caption{Resonance trapping condition in the $\Theta_0 - |d\eta/d\tau|$
  parameter space. Blue crosses: the system will be trapped with certainty. Red
  saltires: the system cannot be trapped. Black circles: the system can
  be trapped for some initial phase $\theta$; trapping is probabilistic in
  this case.}
\label{trap_region}
\end{figure}

Consider a system initially at $\eta_0>1$ and $\Theta_0\ll 1$. The
system passes through the resonance as $\eta$ gradually decreases. 
For $|d\eta/d\tau|\ll 1$, the planet will be trapped in
resonance and its eccentricity can grow to large values. Figure
\ref{Trap} depicts an example. Initially, when $\eta>1$, the
trajectory is circulating around the origin at a small eccentricity. As
$\eta$ passes $1$, a separatrix emerges from the origin.  This
separatrix quickly expands and the trajectory falls into the resonant
(libration) zone.  The separatrix continues to expand as
$\eta$ decreases, while the center of the resonance (the stable fixed point)
moves to increasingly higher $\Theta$ value.
The evolution is now fully adiabatic, and the trajectory is advected with the resonance
to high eccentricity, conserving the its phase-space area.
The mean eccentricity of the planet (as a function of $\eta$) is determined by the location of 
the stable fixed point inside the libration zone:
\be
{\bar\Theta}(\eta) = \frac{1-\eta}{2},
\ee
or 
\be\label{TrapEcc}
{\bar e}(\eta)=\sqrt{{2\epsilon D\over C}{\bar\Theta}(\eta)}=\sqrt{{\epsilon D\over C}(1-\eta)}.
\ee
Once the system is captured in resonance, it will stay in resonance
as $\eta$ continues to decrease\footnote{This is because, as $\eta$ 
becomes more negative, (i) the libration period around the fixed point decreases [see 
Eq.~(\ref{dimless_period2})], so the adibaticity condition $|d\eta/d\tau|\ll 1/\tau_{\rm lib}$
remains well satisfied; (ii) The phase-space area of the libration zone increases
while the area of the trajectory remains constant.}, until the eccentricity becomes too large
and the small-$e$ approximation breaks down.
Note that the growth of ${\bar e}(\eta)$ is unbounded as $\eta$ keeps
decreasing, suggesting that the planet will become unstable if the
system stays in resonance long enough.

For larger values of $|d\eta/d\tau|$, the behavior of the system can
be quite different.  If $|d\eta/d\tau|$ is sufficiently large, the
system passes through the resonance so fast that the eccentricity has
little time to grow before $\eta$ falls below $-1$ and the trajectory
ends up in the inner circulation zone, as is shown in
Fig.~\ref{NotTrap}. For a range of intermediate $|d\eta/d\tau|$
(this range depends on $\Theta_0$; see below), whether a system ends up trapped
in resonance and follows the fixed point inside the libration zone, or
remains circulating around $e=0$, depends on the phase $\theta$ when
it enters resonance. Since this phase is random for realistic systems,
the trapping should be considered probabilistic in this case.
Figure \ref{ProbTrap} depicts an example of this ``mixed'' (probabilistic) behavior.
Note that a larger $\Theta_0$ is chosen for this figure in order to better
visualize both evolutionary trajectories (``trapped'' vs ``non-trapped'').

We have carried out numerical calculations for a wide range of
$d\eta/d\tau$ and $\Theta_0$ values to determine the boundary between
the three behaviors (trapping for certain, probabilistic trapping,
trapping impossible) discussed above. The result is shown in
Fig~\ref{trap_region}.  Two trends are of interest here: First, for
$\Theta_0\ll 1$, the $|d\eta/d\tau|$ values at the boundaries decrease
as $\Theta_0$ decreases, suggesting that slower change of $\eta$ is
required to trap a system with smaller $\Theta_0$. Second, around
$\Theta_0\sim 1$, the probabilistic region quickly expands, and for larger
$\Theta_0$ trapping becomes 
probabilistic even for small $|d\eta/d\tau|$.  
The second trend can be understood using the result
from Borderies \& Goldreich (1984), which shows that trapping become
probabilistic for $\Theta_0\gtrsim 1$ in the $|d\eta/d\tau|\to 0$ regime. 
The first trend can be explained by noting that reaching adiabatic
evolution before the system exits $\eta\in [-1,1)$ is a sufficient
(but not necessary) condition for trapping; this can be used to
approximate the lower boundary of the probabilistic trapping
region.  Specifically, to achieve adiabaticity before leaving the resonance
requires $\Theta$ to reach a large enough value $\Theta_{\rm min}$
such that
\be\label{TrapCon1}
\tau_{\rm lib}(\Theta_{\rm min})\sim -\ln\Theta_{\rm min} 
\lesssim
\left|\frac{d\eta}{d\tau}\right|^{-1}.
\ee
Since $\Theta$ grows exponentially when it is small, we have 
\be\label{TrapCon2}
\ln\Theta_{\rm min} - \ln\Theta_0 \sim \left|\frac{d\eta}{d\tau}\right|^{-1}.
\ee
Substituting this into \eqref{TrapCon1} gives
\be
\left|\frac{d\eta}{d\tau}\right| \lesssim \frac{1}{|\ln\Theta_0|}
\label{eq:trapcon}
\ee
as the condition of ``certain'' resonance trapping.
Therefore, for $\Theta_0\ll 1$, the critical $|d\eta/d\tau|$ at which trapping becomes
probabilistic should decrease as $\Theta_0$ decreases. Note that the condition 
(\ref{eq:trapcon}) has the same scaling as Eq.~(\ref{conditon_on_init}) for the case of
resonance passage, although the actual numerical results are different
(see Figs.~3 and 7).

\subsection{Timescales and Criteria for Eccentricity Excitation and Resonance Capture}

The results of Sections 3.1-3.2 show that significant eccentricity excitation 
(i.e., $e_{\rm f}\gtrsim e_{\rm f,max}$ in the case of increasing $\eta$ or resonance
capture in the case of decreasing $\eta$) requires $|d\eta/d\tau|\lesssim h^{-1}$, or
\eq{\label{general_trap_con}
\left|\frac{d\eta}{dt}\right|^{-1} \gtrsim h\, T_{\rm res},
}
where [see Eqs.~(\ref{eq:tau})-(\ref{eq:tk})]
\eq{\label{T_res_def}
T_{\rm res} = \frac{na^2}{2 |D|\Phi_{B0}}={T_K\over 2|D|}.  
}
The dimensionless quantity $h$ ranges from 1 to 100, depending on the initial
eccentricity; see Figs.~\ref{Ecc_vs_deta} and \ref{trap_region}.

In various applications, the change of $\eta$ can result from the change one of the physical 
parameters of the systems, $X=\{a,a_b,e_b \}$ [see Eq.~(\ref{eq:eta2})].
Since $|\partial\eta/\partial\ln X|\sim \epsilon^{-1}$, we find that, in order to
have significant eccentricity excitation, the timescale for the variation of $X$ must 
satisfy 
\be
T_X\equiv \left|\frac{d\ln X}{dt}\right|^{-1} \gtrsim {h\over\epsilon}T_{\rm res}\equiv 
T_{\rm min}.
\label{eq:tmin3}\ee
Obviously, this criterion is approximate. When $T_X$ is comparable to $T_{\rm min}$, 
numerical calculations are needed to determine the exact behavior of the system near
resonance.

\section{Applications to Circumbinary Planets}

\begin{figure}
\centering
\includegraphics[width=8cm]{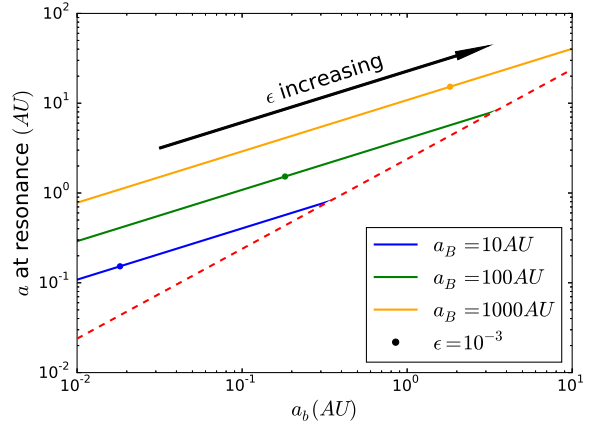}
\caption{Semimajor axis $a$ of planet at the evection resonance as a
  function of $a_b$ for different $a_B$ values. All orbits are assumed
  to be circular, and $M_1=M_2=0.5M_\odot, M_B=1M_\odot$. The dashed
  red line marks the inner boundary of the stability zone (see
    Eq.~\ref{eq:stability}); the outer stability boundary is not
    shown because the lines for the resonance location always
    intersect the inner boundary first.  The filled circles mark the
  location where $\epsilon=10^{-3}$ at resonance.  We see that planets
  at resonance can be stable for a large range of realistic
  $a_b,\,a_B$ values, and $\epsilon\ge 10^{-3}$ for some of the stable
  systems, allowing for appreciable eccentricity excitation even when
  there is no resonance trapping.}
\label{resonant_a}
\end{figure}

We now apply the results of previous sections to examine the effect of
evection resonance on the evolution and stability of circumbinary
planet in a stellar triple. The planet is unlikely to form at the
resonant location due to the small width of resonance; however, it can
be driven into the resonance due to secular effects such as planet
migration and orbital decay of the inner binary.  Passing through the
resonance can make the planet's orbit eccentric. In particular, 
eccentricity can be large if the planet is trapped in the
resonance. This could lead to instability and
ejection/destruction of the planet \citep{MW06}.

For a hierarchical triple system with circumbinary planet, the
semi-major axis of the planet is restricted due to dynamical instability.
\citet{HW99} provided an empirical stability criterion: When all
orbits are circular and coplanar, the planet's orbit is stable only if
\eq{\label{eq:stability}
a_b (1.60+4.12 q_b-5.09 q_b^2)\lesssim a\lesssim a_B (0.46-0.38 q_B), }
where $q_b=M_2/M_{b}$ and $q_B = M_B/M_{\rm tot}$.
Eccentric binary or planet orbits tend to make the stability zone narrower.
Other empirical stability criteria are available (e.g., \citealt{MA01}),
but we will use Eq.~({\ref{eq:stability}) as a guide.

For given binary separations $a_b$ and $a_B$, the evection resonance occurs at the
planetary semi-major axis $a=a_{\rm res}$, given by [see Eq.~(\ref{eq:res2})]
\eq{\label{ap_full}
a_{\rm res}\simeq 0.92\,\left(1+{3e_b^2\over 2}\right)^{\!2/7}\!\!
\left({\mu_{b}^2\over M_{b}M_{\rm tot}}\right)^{\!1/7}\!\!a_b^{4/7}a_B^{3/7}.}
Figure \ref{resonant_a} shows the resonant planet location $a_{\rm res}$ 
as a function of $a_b$ for for different values of $a_B$.
We see that $a_{\rm res}$ falls inside the
stable region for a wide range of realistic $a_b$ and $a_B$ values.

As shown in Section 3.1, the maximum eccentricity the planet can attain
in a resonance passage (without trapping) is of order $\sqrt{\epsilon}$. Evaluating 
Eq.~(\ref{epsilon_def}) at $a=a_{\rm res}$, we find
\be
\epsilon_{\rm res}\simeq 0.66\,\left(1+{3e_b^2\over 2}\right)^{\!\!10/7}\!
{M_B\mu_{b}^{\!3/7}\over (M_{b}M_{\rm tot})^{5/7}}\left({a_b\over a_B}
\right)^{\! 6/7}.
\label{eq:eres}\ee
(In the case of resonance trapping, a higher eccentricity can be
achieved in principle if $\eta$ continues to decreases; see Section 3.2).
Clearly, modest values of $\epsilon_{\rm res}$ (and thus $e_{\rm max}$ in a resonance passage)
can be realized as long as $a_B$ is not too much larger than $a_b$.

As discussed in Section 3.3, in order to achieve resonance capture (when $\eta$ decreases) or
significant eccentricity excitation in a resonance passage (when $\eta$ increases), the 
timescale for the variation of a relevant system parameter $X=\{a,a_b,e_b \}$ 
must be longer than $T_{\rm min}$ 
[see Eq.~(\ref{eq:tmin3})], i.e.,
\be
T_X\gtrsim T_{\rm min}={h\over 2|D|}\left({M_{b}\mu_{b}\over M_B^2}\right)
\left({a_b^2a_B^6\over a^8}\right){1\over n}.
\ee
Setting $a=a_{\rm res}$, we have
\ba
&&T_{\rm min}\simeq  0.213\,{h\over |D|}
\left(1+{3e_b^2\over 2}\right)^{\!\!-13/7}
{M_{b}^{10/7}M_{\rm tot}^{13/14}M_\odot^{1/2}\over
\mu_{b}^{6/7}M_B^2}\nonumber\\
&&\qquad ~~\times \left(\frac{a_b}{1\text{AU}}\right)^{\!\!-12/7}
\!\left(\frac{a_B}{100\text{AU}}\right)^{\!45/14}\text{Myr}.
\label{eq:tmin1}\ea

\subsection{Eccentricity Excitation During Inward Planet Migration}

We first consider the effect of planet migration, with $\dot a\equiv da/dt<0$.
This could result from planet interaction with gas discs
(e.g. \citealt{GT80}; \citealt{Baruteau14})
or with planetesimal discs (e.g. \citealt{HM99}; \citealt{Levison07}).
The migration timescale $T_{\rm mig}=|a/\dot a|$ is highly uncertain, 
ranging from $<1$~Myrs to $>10$~Myrs.

Equation (\ref{eq:eta2}) shows that 
\be
\eta\sim {1\over\epsilon}\left(1-{n_B\over\dot\varpi}\right).
\label{eq:eta3}\ee
With $\dot\varpi\sim (\mu_{b}/M_{b})(a_b/a)^2n\propto a^{-7/2}$
[see Eq.~(\ref{eq:dotomega})], 
we see that $\dot a<0$ leads to $\dot\eta=d\eta/dt>0$. Thus, the planet may 
experience eccentricity excitation without being captured into the resonance
(Section 3.1). The maximum ``final'' eccentricity that can be attained is given by
Eq.~(\ref{THfmax}). Adopting a cononical set of parameters, $M_1=M_2=0.5M_\odot$,
$M_B=1\,M_\odot$ and $e_b=I_B=0$, we find
\be
e_{\rm f,max}= \sqrt{5\epsilon_{\rm res}}=1.05\left({a_b\over a_B}\right)^{\!3/7}.
\label{eq:efmax}
\ee
The resonant planet semi-major axis is 
\be
a_{\rm res}=0.56\, a_b^{4/7}a_B^{3/7}.
\label{eq:ares}
\ee
The mimimum migration timescale required to achieve $e_{\rm f}\sim e_{\rm f,max}/2$ is
\be
T_{\rm min}=3.55\left({h\over 10}\right)\left(\frac{a_b}{1\text{AU}}\right)^{\!\!-12/7}
\!\left(\frac{a_B}{100\text{AU}}\right)^{\!45/14}\text{Myr},
\label{eq:tmin2}\ee
where $h\sim 10$ (see Fig.~\ref{Ecc_vs_deta}).

Figure \ref{planet_mig} shows the $a_b$-$a_B$ parameter space where
eccentricity excitation due to resonance passage is possible. We see
that an eccentricity $\gtrsim 0.2$ can be easily achieved for planets
around short-period binaries ($a_b\lesssim 0.3$~AU). This requires
that the external perturber is relatively close so that $a_{\rm res}$
lies not too far from the instability limit and that the migration
timescale is sufficiently large.

\begin{figure}
\centering
\includegraphics[width=8cm]{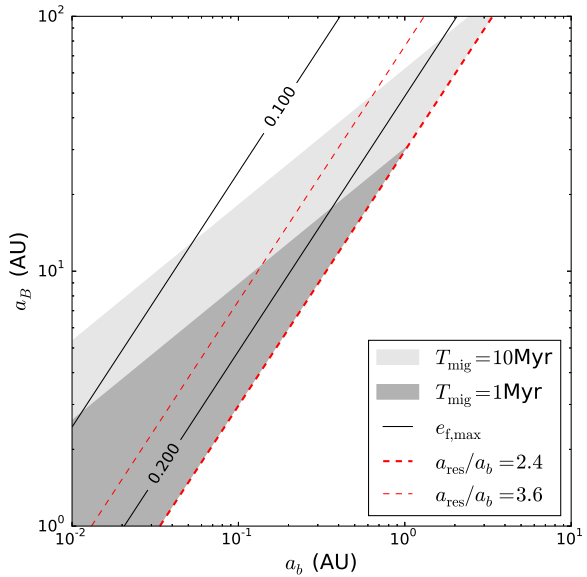}
\caption{Parameter space (in terms of $a_b$ and $a_B$, the semimajor
  axes of the inner and outer binaries) for which significant
  eccentricity excitation occurs during inward planet migration.  The
  system has $M_1=M_2=0.5M_\odot$, $e_b=0$, $M_B=1M_\odot$ and
  $I_B=0$.  For given $a_B$ and $a_b$, the planet's semimajor axis $a$
  is taken as its value at resonance ($a_{\rm res}$).  The black solid
  lines are level curves of constant $e_{\rm f,max}$ (the maximum
  eccentricity that can be attained due to resonance passage; see Eq.~\ref{eq:efmax}).
  In the dark (light) grey region, the planet can be excited to $e_{\rm
    f}\sim e_{\rm f,max}/2$ if the migration timescale ($T_{mig}$) is
  $1$~Myrs ($10$~Myrs) -- This is determined by the condition 
$T_{\rm mig}>T_{\rm min}$ (see Eq.~\ref{eq:tmin2}) with $h=10$.
The lower red-dashed line indicates the inner
  stability boundary $a_{\rm res}/a_b\simeq 2.4$ (see Eq.~\ref{eq:stability}).
}
\label{planet_mig}
\end{figure}

Another factor that needs to be considered is eccentricity damping.
While the planet migrates in the disc, its eccentricity may be damped
by planet-disc interaction. Typically, the eccentricity damping
timescale $T_\text{dmp}$ is much less than $T_\text{mig}$ 
(e.g., \citealt{KN12}).
So for most systems discussed above, we actually have $T_\text{dmp}\lesssim
T_\text{min}\lesssim T_{\rm mig}$.  However, note that for our
systems the timescale of eccentricity libration is $T_\text{res} =
h^{-1}\epsilon T_\text{min} \lesssim 10^{-3} T_\text{min}$ for $h=10$
and $\epsilon \lesssim 10^{-2}$.  This is also the timescale
for eccentricity growth during the resonance passage.
Thus, to achieve eccentricity excitation we only require
$T_\text{res}\lesssim T_\text{dmp}$, which can be satisfied for most 
systems with $T_\text{min}\lesssim T_\text{mig}$. 

We conclude that a circumbinary planet undergoing inward migration can
experience eccentricity excitation when passing through the evection
resonance. If the eccentricity at resonance is sufficiently large, the
planet may suffer instability and be ejected.  On the other hand, if
the eccentricity at resonance is too small to cause instability, then
the planet may survive with a final eccentricity after passing through
the resonance, or its eccentricity may decay to a small value due to
damping from the disc.  The circumbinary planet Kepler-34b has a
significant eccentricity, $e=0.18$ \citep{Welsh12}, whose origin is
unknown. Resonant eccentricity excitation would play a role if an
apprpriate tertiary stellar companion existed during the earlier
planet migration phase.

\subsection{Resonance Trapping Around Shrinking Eccentric Binary}

We consider a planet orbiting around an eccentric inner binary,
which is undergoing orbital decay and circularization due to tidal
dissipation.  Such a decaying eccentric binary represents the final
stage of the ``Lidov-Kozai Shrinkage'' of the inner binary, when the
Lidov-Kozai oscillation (driven by the external binary companion) is
suppressed by various short-range forces and the inner binary
undergoes ``pure'' tidal decay (e.g. \citealt{FT07}; 
\citealt{ML15}; see Section 4.3 below).
Since the planet is strongly coupled to the inner binary [see Eqs.~(6)-(7)],
in the absence of resonance, the secular interaction between the inner
binary and the planet ensures that the planet's orbit remains cricular
and aligned with the inner binary \citep{ML15}.
We now examine how the evection resonance changes the planet's eccentricity.

From Eq.~(\ref{eq:eta2}) or (\ref{eq:eta3}), with
$\dot\varpi\propto (2+3e_b^2) a_b^2$ [see Eq.~(\ref{eq:dotomega})], we see
that as $a_b$ and $e_b$ decrease in time, the parameter $\eta$ also decreases
in time. Thus an out-of-resonance planet could be captured into resonance
as the inner binary shrinks/circularizes (see Section 3.2).

Figures \ref{shrinkin_bin} (for $e_b=0$) and \ref{shrinkin_bin0.8}
(for $e_b=0.8$) show the $a_b-a_B$ parameter space where resonance
capture is certain or has significant probability (see Section 3.2).
We characterize the orbital decay of the inner binary with a constant
timescale $T_{\rm shrink}=|a_b/\dot a_b|$. 
Shrinking due to tidal dissipation tends to be slow, and many systems have $T_{\rm shrink}$ as large as a few Gyr.
We see that for $T_{\rm shrink}
\go 100$~Myr, the binary decay is sufficiently slow that
resonance capture for the planet is likely for a wide range of $a_b$'s
and $a_B$'s.

\begin{figure}
\centering
\includegraphics[width=8cm]{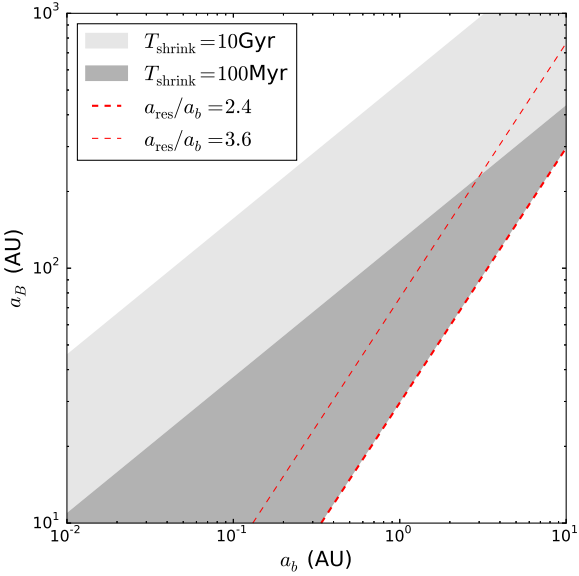}
\caption{Similar to Fig.~\ref{planet_mig}, except for the case of
 shrinking inner binary (with $e_b=0$ and $\dot a_b<0$).
The lighter (dark) grey region indicates the parameter space where
the planet can be trapped in the evection resonance when the 
the characteristic timescale of decrease in $a_b$ is
$T_{\rm shrink}=10$~Gyr ($100$~Myr) -- This is determined by the condition 
$T_{\rm shrink}>T_{\rm min}$ (see Eq.~\ref{eq:tmin2}) with $h=10$.
The lower red-dashed line indicates the inner stability boundary for the planet.
}
\label{shrinkin_bin}
\end{figure}

\begin{figure}
\centering
\includegraphics[width=8cm]{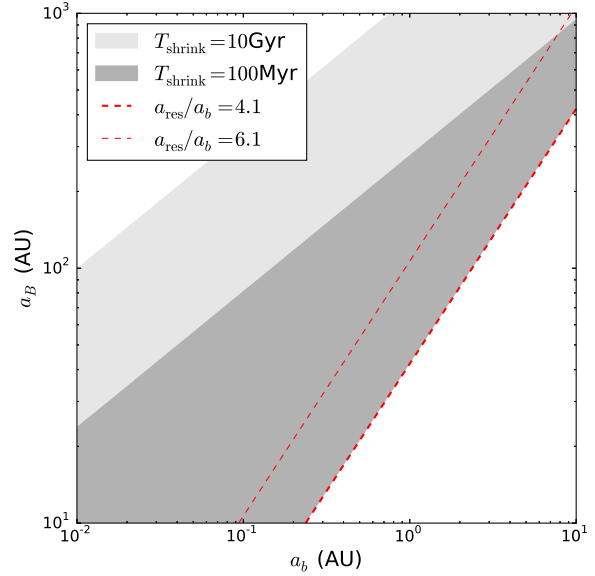}
\caption{Same as Fig.~\ref{shrinkin_bin}, except for $e_b=0.8$. Note that the stability boundary also changes.
}
\label{shrinkin_bin0.8}
\end{figure}

For given semi-major axes of the planet and outer binary, the resonance 
occurs at the inner binary separation (see Eq.~\ref{ap_full})
\ba
&& a_{b,{\rm res}}\simeq 0.21\,\left(1+{3e_b^2\over 2}\right)^{\!-1/2}\!
\left({M_bM_{\rm tot}\over 32\mu_b^2}\right)^{\!1/4}\nonumber\\
&&\qquad\quad \times \left({a\over 1\,{\rm AU}}\right)^{\!7/4}\!\left({a_B\over 30\,{\rm AU}}
\right)^{\!\!-3/4}.
\ea
Given the significant eccentricity excitation of the planet associated with resonance capture, 
the survival of the planet would likely require the initial binary $a_b$ to be less than
$a_{b,{\rm res}}$, so that the resonance can be avoided.

\subsection{Resonance Trapping During Lidov-Kozai Oscillation}
\begin{figure}
\centering
\includegraphics[width=8.3cm]{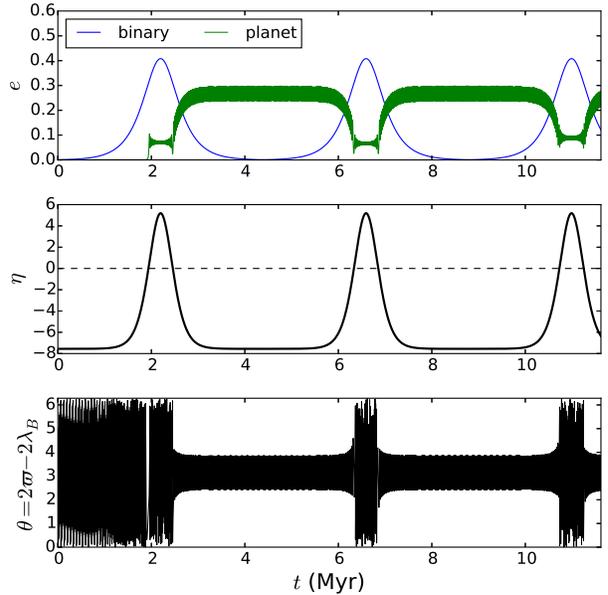}
\caption{Evolution of the planetary eccentricity vector as the inner
  binary undergoes LK oscillations.  The top panel shows the planet's
  eccentricity together with the eccentricity of the inner binary, the
  middle panel shows the evolution of the resonance parameter $\eta$,
  and the bottom panel shows the evection resonance angle $\theta$.
  The system parameters are $M_1=M_2=0.5M_\odot,\, M_B=1M_\odot, a_b =
  1~\text{AU}, a_B =~100\text{AU}$ and $a= 4.2~\text{AU}$.
  The initial inner-outer binary inclination is $I_{B,0}=45^\circ$, which
  gives a maximum inner binary eccentricity of $e_{b,\text{max}}=0.408$.
The planet is trapped in resonance when the eccentricity of the inner binary 
is low, and is out of resonance when $e_b$ is high.
}
\label{Kozai_schematic}
\end{figure}

We now study the evolution of the planet's orbit when the inner binary undergoes
Lidov-Kozai (LK) oscillation in eccentricity and inclination driven by the external 
binary. This problem has been studied by \cite{ML15} using secular theory
(see also \citealt{Martin15}, \citealt{Hamers16}),
without including the effect of the evection resonance. Because of the
strong coupling between the planet and the inner binary, we can assume that the orbital
axes of the planet and the inner binary are always aligned (see \citep{ML15}).
For simplicity, here we neglect tidal dissipation and short-range forces in the 
inner binary, and we also assume that the planet's mass is sufficiently small 
and thus does not influence the LK oscillation of the inner binary.

With these simplifying assumptions, the inner binary undergoes ``pure'' LK oscillation in
$e_b$ and $I_B$ when the initial inclination angle $I_{B,0}$ is greater than 
$\cos^{-1}\sqrt{3/5}\simeq 39.2^\circ$.
The oscillation period is of order $T_{\rm K,b}$, given by
(recall that $n_b$ is the mean motion of the inner binary)
\be
T_\text{K,b} = {n_b a_b^2\over {\mathcal G} M_Ba_b^2/a_B^3}=
{M_{b}\over M_B}\left({a_B\over a_b}\right)^3{1\over n_b}.
\label{eq:T_kb}\ee
In each LK cycle, the inclination oscillates between $I_{B,0}$
and $I_{B,{\rm max}}=39.2^\circ$, while the eccentricity oscillates between $e_{b,0}\simeq 0$
and $e_{b,{\rm max}}$, with $(1-e_b^2)^{1/2}\cos I_B$ conserved.
The maximum inner binary eccentricity is given by 
\be
e_{b,\text{max}} = \sqrt{1-{5\over 3}\cos^2 I_{B,0}}.
\ee
[See \cite{FT07} and \cite{LML15} for a detailed discussion of how various 
short-range forces affect the LK oscillation.]
Note that $I_{B,{\rm max}}$ is the inclination at $e_{b,\text{max}}$ and $I_{B,{\rm max}}<I_{B,0}$.

As $e_b$ oscillates in the LK cycle, the apsidal precession rate of the planet also oscillates between 
$\dot\varpi_0$ and $\dot\varpi_{\rm max}$, given by [see Eq.~(\ref{eq:dotomega})]
\be
\dot\varpi_0={3\Phi_{b0}\over 4a^2n},\qquad
\dot\varpi_{\rm max}=\dot\varpi_0 \left(1+{3e_{b,{\rm max}}^2\over 2}\right).
\ee
Thus, the planet will encounter the evection resonance during the LK cycle if
$\dot\varpi_0<n_B<\dot\varpi_{\rm max}$, or
\eq{\label{eq:condition}
1<{n_B\over \dot\varpi_0}<1+{3e_{b,\text{max}}^2\over 2}.
}
This implies that, for given parameters for the stellar triple ($a_b$, $a_B$ and stellar masses),
resonance encounter occurs only for a restricted range of planetary semimajor axis, i.e.,
\be
1<{a\over a_{\rm res,0}}<\left(1+{3e_{b,{\rm max}}^2\over 2}\right)^{\!2/7},
\ee
where $a_{\rm res,0}$ is given by Eq.~(\ref{ap_full}) with $e_b=0$.
For a given $a$ in this range, 
the resonance occurs at the inner binary eccentricity $e_{b,{\rm res}}$, given by 
\be
{n_B\over \dot\varpi_0}=1+{3e_{b,\text{res}}^2\over 2}.
\ee

When the condition (\ref{eq:condition}) is satisfied, the planet will
pass the resonance twice in each LK cycle: 
(i) During the increasing-$e_b$ phase, $\eta$ changes from from
$\eta_0\equiv \eta(e_b=0)<0$ to $\eta_{\rm max}\equiv\eta(e_b=e_{b,\text{max}})>0$;
this can excite the planet's eccentricity (if
the initial eccentricity is small) but does not lead to resonance capture.
(ii) During the decreasing-$e_b$ phase, $e_b$ passes the
resonance for the second time, and $\eta$ decreases from $\eta_{\rm max}>0$ 
to $\eta_0<0$; this can lead to resonance capture.
From Eq.~(\ref{eq:eta2}), we find
\be
\eta_0\simeq {4\over 5(1+\cos I_{B,0})^2\epsilon}\left(1-{n_B\over 
\dot\varpi_0}\right)=-{6 e_{b,\text{res}}^2
\over 5(1+\cos I_{B,0})^2\epsilon},
\ee
and 
\be
\eta_{\rm max}\simeq {6(e_{b,\text{max}}^2-e_{b,\text{res}}^2)
\over 5(1+\cos I_{B,\text{max}})^2\epsilon}.
\ee

Figure \ref{Kozai_schematic} shows an example of the evolution of the planetary 
eccentricity as the inner binary undergoes LK oscillation. We see that, except for the
initial phase of the calculation, 
the planet spends most of the time in the trapped resonance (librating) state
during the low-$e_b$ phase of the LK oscilation. 
It is thrown out the resonance only during the brief high-$e_b$ phase, as the system crosses
$\eta=1$ and the area of the libration zone shrinks to zero.
The mean eccentrcity of the planet in the trapped state, ${\bar e}_{\rm trap}$,
can be estimated from Eq.~(\ref{TrapEcc}), where $\eta$ is evaluated at $e_b=0$. 
We then have ${\bar e}_{\rm trap}={\bar e}(\eta=\eta_0)$ given by 
\be
{\bar e}_{\rm trap}\simeq \left[ \frac{3}{4}e_{b,\text{res}}^2
+{5\epsilon\over 8}(1+\cos I_{B,0})^2\right]^{1/2}
\simeq {\sqrt{3}\over 2}\,e_{b,\text{res}}.
\ee

\begin{figure}
\centering
\includegraphics[width=8cm]{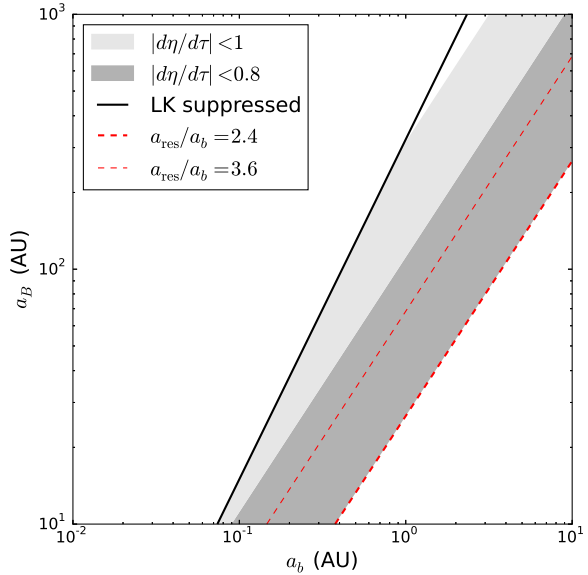}
\caption{Parameter space (in terms of $a_b$ and $a_B$, the semi-major axes of the
inner and outer binaries) for which resonance trapping can happen when the inner
binary undergoes LK oscillations. 
Dark grey: $|\dot\eta|<0.6$. Light grey: $|\dot\eta|<1$. Thick red line:
$a_{\rm res}/a_b=2.4$ (the stability limit).
Thick black line: 
$T_{\rm GR}=T_{\rm K,b}$ (above this line, the short-range force due to GR
can fully suppress LK oscillations of the inner binary).
We see that $|d\eta/d\tau|\sim 1$ in most of the allowed parameter space.
Our analysis (see text) shows that the system can be trapped in resonance in most cases.
}
\label{KOZAI}
\end{figure}

As discussed at the beginning of Section 4, resonance capture requires
that the LK timescale of the inner binary $T_{\rm K,b}$ (see
Eq.~\ref{eq:T_kb}) be not much smaller than $T_{\rm min}$ (see
Eq.~\ref{eq:tmin1}). For systems of interest here, we find $T_{\rm
  K,b}$ is of the same order as $T_{\rm min}$. More precisely, in
light of the result of Section 3.2 (especially
Fig.~\ref{trap_region}), we can evaluate $d\eta/d\tau$ at resonance
using the exact equations for the $e_b$ and $I_{B}$ oscillations of
the inner binary.
Figure \ref{KOZAI} shows the parameter space (in terms of $a_b$ and $a_B$)
where $|d\eta/d\tau|<1$ can be realized.
Note that in addition to the stabililty requirement for the planet,
we also restrict to the $a_b$-$a_B$ space where LK osicllations are not suppressed
by General Relativity (GR). This requres that $T_{\rm K,b}$ be shorter than
the GR-induced apsidal precession timescale of the inner binary, i.e.,
\be
T_{\rm K,b}\lesssim T_{\rm GR}={a c^2\over 3n_b\mathcal GM_b}.
\ee
In the example depicted in Fig.~\ref{Kozai_schematic}, 
the planet attains a significant eccentricity in the first ($\eta$-increasing) 
resonance passage (during the first half of the LK cycle). The $\Theta$ value when the planet 
crosses the resonance again (during the $\eta$-decreaisng phase) can be $\sim 1$.
From Fig.~\ref{trap_region} we see that the planet can be
trapped into resonance for $|\dot\eta|\sim 1$ if $\Theta$ is large
enough when the system enters resonance. This explains why the system can be
trapped in resonance despite having $|d\eta/d \tau|\sim 1$.

\subsection{N-Body Calculations}

To test the validity of the results discussed in Section 4.3,
we have performed a N-body integration using the \textit{Mercury} code
\citep{CHAMBERS99}.
The result is shown in Figure \ref{KOZAI_num}. The parameters are the
same as the system depicted in Fig.~\ref{Kozai_schematic}, with
initial inner binary eccentricity $e_b=e=10^{-3}$. The planet and
binary eccentricities are evaluated by calculating the eccentricity
vectors (Laplace-Runge-Lenz vectors) of the bodies.  Due to the
perturbation from the inner binary, the initial eccentricity of the
planet exhibits small oscillations (with an amplitude $\sim 10^{-2}$)
even before any resonance encounter; this does not occur in the
secular approximation, when the planet and the inner binary are
orbit-averaged.  Because of this forced initial eccentricity, the
excited eccentricity of the planet during the increasing-$\eta$
resonant passage is larger than the ``secular'' result shown in
Fig.~\ref{Kozai_schematic}. The eccentricity libration amplitude when
the system is captured in resonance is also slightly larger due to the
increased phase space area of the trajectory. Besides these small
differences, our N-body result agrees well with the result obtained
from the integration of orbit-averaged equations as depicted in
Fig.~\ref{Kozai_schematic}. During the third LK cycle of the
binary, the planet fails to be captured into resonance as $\eta$
decreases below 0 (see Fig.~\ref{KOZAI_num}), showing the
probabilistic nature of trapping in this regime. Note that for
Fig.~\ref{Kozai_schematic} trapping is also probabilistic; the absence
of non-trapping passage is a coincidence.

We see from Fig.~\ref{KOZAI_num}) that, during the first $\sim 10$~Myr
of the integration (the first three-four LK cycles), the planet is
stable although both $e_b$ and $e$ can be significant. This stability
arises because when $e$ reaches the maximum value due to resonance
trapping, $e_b$ is always near its minimum. However, the planet
becomes unstable when it passes resonance with increasing $\eta$ for
the fourth time; this time the eccentricity excitation happens to be
slightly larger, because the system fails to be trapped in the
previous decreasing $\eta$ passage\footnote{When the system is trapped
  as $\eta$ decreases past 0, the phase space area is conserved and
  the planet's eccentricity when $e_b>e_{\rm b,res}$
(corresponding to positive $\eta$) will be the same as that in the previous
  cycle. However, if it is not trapped, then the phase space area may
  change and the eccentricity when the system next enters the $e_b>e_{\rm b,res}$ can be different. 
  In the example depicted in Fig.~\ref{KOZAI_num}, the phase space area
  increases, leading to a higher planetary eccentricity.}.  This slightly
larger $e$, together with the large $e_b$, makes the system unstable
and the planet is ejected.

\begin{figure}
\centering
\includegraphics[width=8cm]{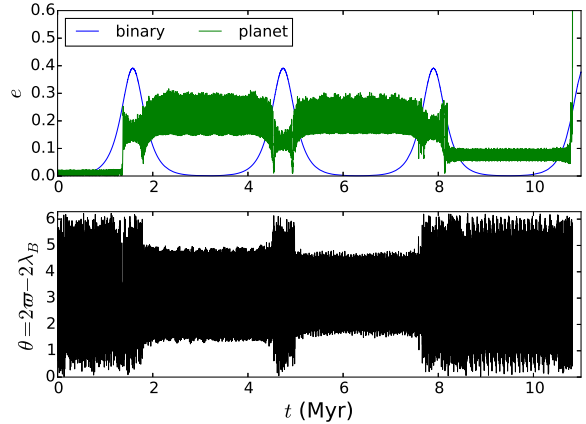}
\caption{N-body integration for the system depicted in Fig.~\ref{Kozai_schematic}. The planet 
attains a large eccentricty at $t\sim 11$~Myr and is ejected.
}
\label{KOZAI_num}
\end{figure}

\section{Application to Multiplanet Systems}

We now apply the theory of evection resonance (Sections 3 and 4) to
multiplanet systems with external binary companions. This problem has
been studied recently by \citep{TS15}. Our goal here is to
adapt the results of previous sections to determine the explicit
conditions for efficient excitation of planetary eccentricities.

Consider a wide binary where the primary hosts two planets, one of
them much more massive than the other.  Let the primary, the binary
companion and the massive planet have mass $M_\star$, $M_B$ and $m_p$
respectively ($m_p\ll M_\star$); let the massive planet have
semi-major axis $a_p$ and zero eccentricity, and the small planet,
considered as a test mass, have semi-major axis $a$ and eccentricity
$e$. As in the circumbinary planet problem, we assume that the planets
are coplanar, while the outer binary can have an arbitrary inclination
angle $I_B$ relative to the planetary orbits.  The small planet
undergoes apsidal precession caused by the quadrupole moment of the
massive planet. Evection resonance occurs when apsidal precession
frequency ($\dot\varpi$) equals $n_B$, the mean motion of the
binary. During migration, $\dot\varpi$ changes due to the evolution of
the planet's semi-major axes, and the system may pass through evection
resonance.

\subsection{Inner Massive Planet}

First consider the case when the inner planet is massive ($a_p<a$). In
this regime we can directly apply the previous results for
circumbinary planet by changing $(M_1,M_2,a_b,e_b)$ to
$(M_\star,m_p,a_p,e_p)$ and taking $e_p = 0$. We use $\Phi_{p}$ to
denote the quadrupole perturbation potential of the massive planet
(see Eq.~\ref{eq:Phib}) and define $\Phi_{p0}$ analogously,
i.e. $\Phi_{b0}={\mathcal G} m_p a_p^2/a^3$.  This is a rather
approximate model, since we expand in $a_p/a$ and only keep the first
nontrivial term in the perturbing potential.  However, this model
should be adequate to capture the qualitative behavior and scalings of
the system.

To model the effect of planet migration we assume that the massive
planet has a fixed semi-major axis $a_p$ (and eccentricity
$e_p=0$)\footnote{We neglect any possible evolution of the orbit of
  the massive planet. For exaple, when $I_B$ is sufficiently large,
  the Lidov-Kozai oscillations of the massive planet induced by the
  binary companion may be suppressed by planet-planet secular
  interactions.}, while the outer (small) planet migrates inward or
outward with a constant rate $|{\dot a}/a| \equiv T_{\rm mig}^{-1}$.
Since the apsidal precession rate $\dot\varpi\propto a^{-7/2}$, we
have $\partial\eta/\partial a<0$ (see Eq.~\ref{eq:eta2}).
Thus, for slow migration, $\eta$
increases as the outer planet migrates inward, leading to resonance
passage and eccentricity excitation (see Section 3.1), and $\eta$ decreases as the
planet migrates outward, leading to resonance trapping (Section 3.2).

The typical timescale of migration driven by gas in the disc
can range from $<1$~Myrs to $>10$~Myrs, while outward migration due
to scatterings with planetesimals occurs on longer timescales (100's Myrs).
The result of Section 3 (see Eq.~\ref{eq:tmin3} or Eq.~\ref{eq:tmin1}) shows that to ensure
eccentricity excitation or resonance capture the migration time must satisfy (with $e_p=0$)
\ba
&&T_{\rm mig}\go T_{\rm min}\simeq 767\,{h\over |D|}
\left({m_p\over M_J}\right)^{\!\!-6/7}
\left({M_\star^{20}M_{\rm tot}^{13}\over M_B^{28}M_\odot^5}\right)^{\!\!1/14}\nonumber\\
&&\qquad\qquad\qquad \times \left(\frac{a_p}{1\text{AU}}\right)^{\!\!-12/7}
\!\left(\frac{a_B}{200\text{AU}}\right)^{\!45/14}\text{Myr},
\label{eq:tmin4}\ea
where $M_{\rm tot}=M_\star+M_B$, and 
we have set $a=a_{\rm res}$ (the resonance semi-major axis of the planet), with
\be
a_{\rm res}\simeq 1.2\left({m_p\over 1\,M_J}\right)^{\!\!2/7}
\!\left({M_\odot^2\over M_\star M_{\rm tot}}\right)^{\!\!1/7}
\!\left({a_p\over 1\,{\rm AU}}\right)^{\!4/7}\!\left({a_B\over 200\,{\rm AU}}\right)^{\!3/7}{\rm AU}.
\ee
Figure \ref{multi_inner} shows the region in $a_p$-$a_B$ parameter space
where resonance trapping (for outward migration) or eccentricity excitation
comparable to $e_{\rm f,max}$ (for inward migration) is possible.
From Eq.~(\ref{THfmax}) we find that the maximum eccentrcity that can be achieved 
in a resonance passage is given by (see also Eq.~\ref{eq:eres})
\ba
&& e_{\rm f,max}=\sqrt{2\epsilon_{\rm res} D\over C}\nonumber\\
&&\quad \simeq 0.042\left({m_p\over 1\,M_J}\right)^{\!3/14}\!
{M_B^{1/2} M_\odot^{3/14}\over (M_\star M_{\rm tot})^{5/14}}
\left({200a_p\over a_B}\right)^{\!3/7},
\ea
where we have used $2D/C=5$ (for $e_p=I_B=0$).
We see that for inward migration, only modest eccentricity ($e_{\rm f,max}\lesssim 0.1$) 
can be attained in a resonance passage.  The
reason is that $\epsilon$ is quite small for any binary companion that
induces resonance on the (small) planet which satisfies
the stability criterion.
For outward migration, the region in the parameter space where the system
can be trapped in resonance is much larger, since the timescale can be
larger for outward migration driven by planetesimal scatterings. Also
note that in this case, the eccentricity excitation for a system
trapped in resonance is not limited by the small value of $\epsilon$
(see Eq.~\ref{TrapEcc}).

Note $T_{\rm min}$ depends on the planet mass.  Figure
\ref{multi_inner} uses $m_p= 10\,M_J$ and solar-mass stars. For
smaller planet mass, the region allowing resonance trapping
shrinks. For example, at $m_p=1\, M_J$, no system with $T_{\rm
  mig}\lesssim 250$~Myr and $a_p>0.01$~AU allows trapping if we take
$h=10$. Therefore, resonance trapping only occurs for the most massive
planets.

\cite{TS15} studied similar systems in greater detail.
Here we compare their result with ours. The canonical system
considered by \citeauthor{TS15} has $m_p\simeq 10\,M_J$, $a_p=5$~AU,
$a_B=1000$~AU and $T_{\rm mig}\simeq 220$~Myr. As shown in
Fig.~\ref{multi_inner}, this system lies far outside the region in
which we consider trapping likely, and yet Touma \& Sridhar found
resonance trapping in their calculations.  This difference arises because we use $h=10$
in Fig.~15 (i.e. we require $|d\eta/d\tau| < 0.1$ for effective trapping; see
Sections 3.2-3.3), which is appropriate for the (scaled) initial
eccentricity $\Theta_0\ll 1$.  On the other hand, Touma \& Sridhar chose a relatively
large initial eccentricity, $e_0 \simeq 0.05$, corresponding to
$\Theta_0\sim 1$. As discussed in Section 3.2 (see Fig.~7), for 
$\Theta_0\go 1$, the system is in the 
probabilistic trapping regime with nontrivial trapping probability.

\begin{figure}
\centering
\includegraphics[width=8cm]{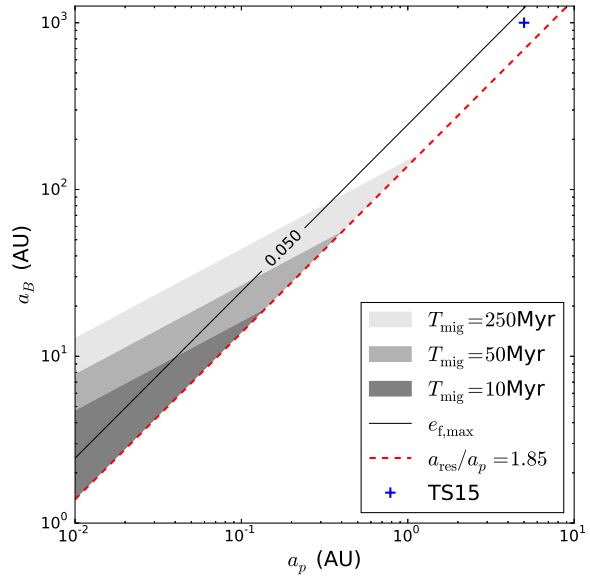}
\caption{Parameter space (in terms of $a_p$ and $a_B$, the semi-major
  axes of the massive planet and the external binary) for which
  significant eccentricity excitation or resonance trapping occurs
  during the migration of a low-mass planet. The system has $M_\star =
  M_B=1M_\odot$, $m_p = 10\,M_{\rm Jup}$ and $I_B=0$. The black solid
  line marks $e_{\rm f,max}=0.05$.  The red-dashed line indicates the
  inner stability boundary $a_{\rm res}/a_p\simeq 1.85$, as given by
  the fitting formula of \protect\cite{PETROVICH15}.  In the three grey regions,
  the small planet can be excited to $e_{\rm f}\sim e_{\rm f,max}/2$
  for inward migration or trapped in resonance for outward migration
  if the migration timescale ($T_{\rm mig}$) is $10$~Myrs, $50$~Myrs
  and $250$~Myrs respectively. This is determined by the condition
  $T_{\rm mig}\go T_{\rm min}$ (see Eq.~\ref{eq:tmin4}) with $h=10$ --
  as appropriate when the initial eccentricity $e_0$ is sufficiently
  small (such that $\Theta_0\ll 1$). The blue cross marks the canonical system
  considered by \protect\cite{TS15}, which has $a_p=5$~AU and $a_B=1000$~AU (see text for discussion).}
\label{multi_inner}
\end{figure}

\subsection{Outer Massive Planet}

Next we consider the case when the massive planet is the outer planet ($a_p>a$). 
The potential from the massive planet ($m_p$) on the inner (small) planet is
\eq{
\Phi_p = -\frac{\Phi_{p0}}8 \left(2+3e^2\right),
}
where
\eq{
\Phi_{p0} = \frac{\mathcal G m_pa^2}{a_p^3}.
}
The dimensionless ratio $\epsilon$ is
\eq{
\epsilon = \frac{\Phi_{B0}}{\Phi_{p0}} = \frac{M_Ba_p^3}{m_pa_B^3}
}
Although $\Phi_p$ has a different dependence on $e$ compared to $\Phi_b$ (see Eq.~2), 
we can use the same procedure as in Section 2.2 to simplify the Hamiltonian. We obtain the 
same dimensionless Hamiltonian as Eq.~(17), but with 
\eal{
A = -\frac{3}{4},~~~C = \frac 3 8 
}
while $B, D$ are the same.  The sign of $C$ differs from Eq.~(20);
thus in order to maintain the form of the scaled Hamiltonian $K$ (Eq.~26) while
keeping $\Theta$ positive, the scaled variables/parameters need also be different:
\eal{
&\theta = -2\varpi+2\lambda_B+\pi ,\\
&\Theta = -\frac C{\epsilon D} \Gamma ,\\
&\eta = -\frac{A+\epsilon B+(n_Bna^2/\Phi_{p0})}{\epsilon D}.
}
We can see that $\Theta\sim \epsilon^{-1} e^2$ still holds. One
interesting property of the new variable $\theta$ (Eq.~77, as compared to 
Eq.~22) is that when a system is trapped in resonance ($\theta=\pi$), 
$\varpi-\lambda_B$ is 0 or $\pi$, i.e., the planet will
be aligned or anti-aligned with the binary when the system is in
resonance. This is in contrast to the situations studied in previous sections, 
where resonance corresponds $\varpi-\lambda_B=\pm \pi/2$.

The new $\eta$ (as a function of $A,B,C,D$) defined in Eq.~(79) differs from Eq.~(27)
by a sign, and can be written as 
\eq{
\eta = \frac{4}{5(1+\cos I_B)^2}\frac{1}{\epsilon}\left[\frac{n_B}{\dot\varpi(e=0)}-1+\frac 1 2
\, \epsilon \,(1-3\cos^2 I_B)\right].
}
Also note that the apsidal precession frequency is given by 
\eq{
\dot\varpi(e=0) = \frac 34\frac{\Phi_{p0}}{na^2} \propto a^{3/2}.
}
Clearly, $\partial \eta/\partial a<0$. Thus, $\eta$ increases as
the small planet migrates inward, leading to eccentricity excitation,
and $\eta$ decreases as the planet migrates outward, leading to resonance trapping. 
The location of resonance, set by $n_B \simeq \dot\varpi (e=0)$, is given by 
\be
a_{\rm res} \simeq \left({16 M_\star M_{\rm tot}\over 9 m_p^2}\right)^{\!1/3}
{a_p^2\over a_B}.
\ee
The corresponding $e_{\rm f,max}$ is
\ba
&& e_{\rm f,max}=\sqrt{2\epsilon |D|\over C}=\sqrt{20\epsilon}\nonumber\\
&&\qquad \simeq 0.14\,\left({M_B\over 10^3m_p}\right)^{\!1/2}
\left({100 a_p\over a_B}\right)^{\!3/2},
\ea
where the second equality assumes $e_p=I_B=0$. The minimum migration timescale for
efficient eccentricity excitation is 
\ba
\label{Tmin_outer}
&&T_{\rm min}={h\over 2|D|}{M_\star m_p\over M_B^2}{a_B^6\over a^3a_p^3}{1\over n}
\nonumber\\
&&\qquad \simeq 54\,{h\over |D|}\!\left(\!{m_p\over M_J}\!\right)^2
{M_\odot^{5/2}\over M_B^2M_{\rm tot}^{1/2}}\nonumber\\
&&\qquad\qquad \times \left({a_B\over 100\,{\rm AU}}\right)^{\!15/2}
\left({a_p\over 1\,{\rm AU}}\right)^{\!-6}{\rm Myr}.
\ea

Figure \ref{multi_outer} shows the parameter space where nontrivial eccentricity 
excitation or resonance trapping occurs.
We see that the qualitative behavior is 
similar to the inner massive planet case discussed in Section 5.1;
the major difference is that the region allowing for significant eccentricity excitation or
resonance trapping now lies almost parallel to the instability limit, and such
difference is due to the difference of the scaling of $T_{\rm min}$. 
Although Eq.~\eqref{Tmin_outer} suggests that $T_{\rm min}$ decreases as
$m_p$ decreases, resonance trapping or significant eccentricity excitation is
still less likely for smaller $m_p$. This is because smaller $m_p$
leads to smaller $a_p/a_{\rm res}$ and the system is more prone to instability.
Moreover, since we require $a_{\rm res}<a_p$, resonance becomes impossible when 
$m_p$ is too small (see Eq.~85).

Together, Figures \ref{multi_inner} and \ref{multi_outer} show that
the region where nontrivial eccentricity excitation occurs occupies a
relatively small portion of the parameter space both inner/outer
massive planet cases.
This suggests that evection resonance plays only a modest role in
most of the multiplanet systems with binary companions.

\begin{figure}
\centering
\includegraphics[width=8cm]{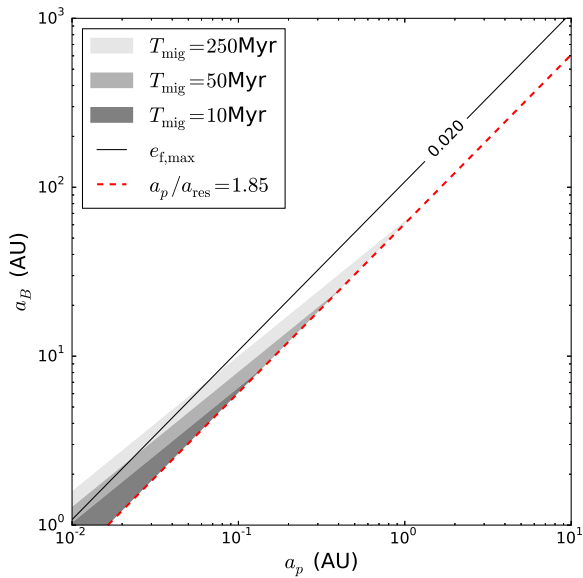}
\caption{Similar to Fig.~\ref{multi_inner} except for the outer
  massive planet case ($a<a_p$). We see that the qualitative behavior
  is similar to the $a>a_p$ case, and the shape of the region for
  eccentricity excitation changes due to the different scaling
  relations.}
\label{multi_outer}
\end{figure}

\section{Summary}

We have developed an analytic theory of evection resonance for
circumbinary planets and multiplanet systems under the perturbation
from an external companion. The resonance occurs when the apsidal
precession of the planet, driven by the quadrupole moment associated
with the inner binary or another massive planet, equals the orbital
frequency of the external perturber. The theory is quite general and
can be applied to various astrophysical/planetary setups. The key
results of our paper are:

1. The dynamics of a planet near the evection resonance is described
by the one-parameter nondimensional Hamiltonian, Eq.~(26). This Hamiltonian 
has the same form as that describing second-order mean-motion resonances.

2. As the parameter $\eta$ (see Eqs.~27-28) changes due to the
long-term evolution of the system (e.g., planet migration or inner
binary shrinkage), the planet may pass through the resonance (for
increasing $\eta$) or experience resonance capture (for decreasing
$\eta$). In the former case, the maximum planetary eccentricity that
can be attained is given by Eq.~(43); in the latter case,
the eccentricity could continue to grow to larger values (as $\eta$ keeps
decreasing) as long as the planet stays in resonance. 
(Resonance escape may occur when the eccentricity becomes ``nonlinear'' and
the planet becomes unstable.)
In both cases, in order to achieve appreciable eccentricity excitation
(comparable to $e_{\rm f,max}$) or resonance capture, the
dimensionless parameter $\eta$ must vary slowly (Eq.~49; see also
Fig.~7), or equivalently, the timescale of variation for the dimensional system
parameter (such as planetary or binary semi-major axis) must be longer
than a minimum value (Eq.~52) which depends on the planet's initial eccentricity.

3. Applying our theory to circumbinary planets with external stellar
perturbers (Section 4), we show that: (i) Inward planetary migration
may lead to resonance passage, producing eccentric planet. An example
of such circumbinary planet is Kepler-34b, with $e=0.18$ \citep{Welsh12}.  
(ii) The shrinkage of inner binary may result in
resonance capture of the planet, potentially leading to its destruction.  
(iii) The planet may periodically enter and exit
evection resonance as the inner binary undergoes Lidov-Kozai
oscillations. Furthermore, our N-body calculations (Section 4.4)
suggest that during this process the planet is likely to become
unstable and be destructed due to its excited eccentricity.
Taken together, our results suggest that survival
of planet around a shrinking binary \citep{ML15} likely
requires that the initial binary semi-major axis to be less than a
critical value, so that the evection resonance can be avoided (see
Eq.~62).

4. Applying our theory to multiplanet systems with external stellar
perturbers (Section 5), we clarify the conditions for significant
eccentricity excitation due to evection resonance. We approximate a
multiplanet system as consisting of a test-mass planet either inside
or outside a massive planet. We find that the conditions for
eccentricity excitation/capture are qualitatively similar for the two
cases, despite the difference in scaling relations for the two
different planetary architectures.
In general, we find that the parameter space where nontrivial
eccentricity excitation occurs is relatively small, suggesting that
evection resonance plays only a modest role in most of the multiplanet
systems with binary companions.

\section*{acknowledgements}
This work has been supported in part by NSF grant AST-1211061, and
NASA grants NNX14AG94G and NNX14AP31G.  WX acknowledges the supports from
the Hunter R. Rawlings III Cornell Presidential Research Scholar Program
and Hopkins Foundation Summer Research Program for undergraduates.


\bibliographystyle{mn2e}
\bibliography{BIB}

\begin{thebibliography}{26}
\expandafter\ifx\csname natexlab\endcsname\relax\def\natexlab#1{#1}\fi

\bibitem[{{Baruteau} {et~al}\mbox{.}(2014){Baruteau}, {Crida}, {Paardekooper},
  {Masset}, {Guilet}, {Bitsch}, {Nelson}, {Kley}, \& {Papaloizou}}]{Baruteau14}
{Baruteau} C. {et~al.}, 2014, Protostars and Planets VI, 667

\bibitem[{Borderies \& Goldreich(1984)}]{BG84}
Borderies N., Goldreich P., 1984, Celestial mechanics, 32, 127

\bibitem[{Chambers(1999)}]{CHAMBERS99}
Chambers J.~E., 1999, MNRAS, 304, 793

\bibitem[{Doyle {et~al}\mbox{.}(2011)Doyle, Carter, Fabrycky, Slawson, Howell,
  Winn, Orosz, Pr?sa, Welsh, Quinn, {et~al.}}]{Doyle11}
Doyle L.~R. {et~al.}, 2011, Science, 333, 1602

\bibitem[{Fabrycky \& Tremaine(2007)}]{FT07}
Fabrycky D., Tremaine S., 2007, ApJ, 669, 1298

\bibitem[{{Goldreich} \& {Tremaine}(1980)}]{GT80}
{Goldreich} P., {Tremaine} S., 1980, ApJ, 241, 425

\bibitem[{Hahn \& Malhotra(1999)}]{HM99}
Hahn J.~M., Malhotra R., 1999, AJ, 117, 3041

\bibitem[{{Hamers}, {Perets} \& {Portegies Zwart}(2016){Hamers}, {Perets}, \&
  {Portegies Zwart}}]{Hamers16}
{Hamers} A.~S., {Perets} H.~B., {Portegies Zwart} S.~F., 2016, MNRAS, 455, 3180

\bibitem[{Holman \& Wiegert(1999)}]{HW99}
Holman M.~J., Wiegert P.~A., 1999, AJ, 117, 621

\bibitem[{{Kley} \& {Nelson}(2012)}]{KN12}
{Kley} W., {Nelson} R.~P., 2012, Annual Review of Astronomy \& Astrophysics,
  50, 211

\bibitem[{{Kostov} {et~al}\mbox{.}(2015){Kostov}, {Orosz}, {Welsh}, {Doyle},
  {Fabrycky}, {Haghighipour}, {Quarles}, {Short}, {Cochran}, {Endl}, {Ford},
  {Gregorio}, {Hinse}, {Isaacson}, {Jenkins}, {Jensen}, {Kull}, {Latham},
  {Lissauer}, {Marcy}, {Mazeh}, {Muller}, {Pepper}, {Quinn}, {Ragozzine},
  {Shporer}, {Steffen}, {Torres}, {Windmiller}, \& {Borucki}}]{Kostov15}
{Kostov} V.~B. {et~al.}, 2015, ArXiv e-prints

\bibitem[{Levison {et~al}\mbox{.}(2007)Levison, Morbidelli, Gomes, \&
  Backman}]{Levison07}
Levison H.~F. {et~al.}, 2007, Protostars and planets V, 1, 669

\bibitem[{{Liu}, {Mu{\~n}oz} \& {Lai}(2015){Liu}, {Mu{\~n}oz}, \&
  {Lai}}]{LML15}
{Liu} B., {Mu{\~n}oz} D.~J., {Lai} D., 2015, MNRAS, 447, 747

\bibitem[{Mardling \& Aarseth(2001)}]{MA01}
Mardling R.~A., Aarseth S.~J., 2001, MNRAS, 321, 398

\bibitem[{Martin, Mazeh \& Fabrycky(2015)Martin, Mazeh, \& Fabrycky}]{Martin15}
Martin D.~V., Mazeh T., Fabrycky D.~C., 2015, MNRAS, 453, 3554

\bibitem[{{Mu{\~n}oz} \& {Lai}(2015)}]{ML15}
{Mu{\~n}oz} D.~J., {Lai} D., 2015, PNAS, 112, 9264

\bibitem[{Mudryk \& Wu(2006)}]{MW06}
Mudryk L.~R., Wu Y., 2006, AJ, 639, 423

\bibitem[{Murray \& Dermott(1999)}]{MD99}
Murray C.~D., Dermott S.~F., 1999, Solar system dynamics, Cambridge university
  press, pp. 321--408

\bibitem[{Peale(1986)}]{PEALE86}
Peale S., 1986, in IAU Colloq. 77: Some Background about Satellites, Vol.~1,
  pp. 159--223

\bibitem[{Petrovich(2015)}]{PETROVICH15}
Petrovich C., 2015, ApJ, 808, 120

\bibitem[{{Spalding}, {Batygin} \& {Adams}(2016){Spalding}, {Batygin}, \&
  {Adams}}]{SBA16}
{Spalding} C., {Batygin} K., {Adams} F.~C., 2016, ApJ, 817, 18

\bibitem[{Touma \& Wisdom(1998)}]{TW98}
Touma J., Wisdom J., 1998, AJ, 115, 1653

\bibitem[{Touma \& Sridhar(2015)}]{TS15}
Touma J.~R., Sridhar S., 2015, Nature, 524, 439

\bibitem[{Tremaine, Touma \& Namouni(2009)Tremaine, Touma, \& Namouni}]{TTN09}
Tremaine S., Touma J., Namouni F., 2009, AJ, 137, 3706

\bibitem[{Tremaine \& Yavetz(2014)}]{TY14}
Tremaine S., Yavetz T.~D., 2014, Am. J. Phys, American Journal of Physics, 769

\bibitem[{{Welsh} {et~al}\mbox{.}(2012){Welsh}, {Orosz}, {Carter}, {Fabrycky},
  {Ford}, {Lissauer}, {Pr{\v s}a}, {Quinn}, {Ragozzine}, {Short}, {Torres},
  {Winn}, {Doyle}, {Barclay}, {Batalha}, {Bloemen}, {Brugamyer}, {Buchhave},
  {Caldwell}, {Caldwell}, {Christiansen}, {Ciardi}, {Cochran}, {Endl},
  {Fortney}, {Gautier}, {Gilliland}, {Haas}, {Hall}, {Holman}, {Howard},
  {Howell}, {Isaacson}, {Jenkins}, {Klaus}, {Latham}, {Li}, {Marcy}, {Mazeh},
  {Quintana}, {Robertson}, {Shporer}, {Steffen}, {Windmiller}, {Koch}, \&
  {Borucki}}]{Welsh12}
{Welsh} W.~F. {et~al.}, 2012, Nature, 481, 475

\end{thebibliography}

\end{document}